\documentclass[prb,twocolumn,showpacs,amssymb,superscriptaddress]{revtex4-2}
\usepackage{graphicx}
\usepackage{amsmath}
\usepackage{amsfonts}
\usepackage{amsthm}
\usepackage{amssymb}
\usepackage{amsbsy}
\usepackage{wasysym}
\usepackage{bm}
\usepackage{bbm}
\usepackage{mathrsfs}
\usepackage{color}
\usepackage{hyperref}
\usepackage{braket}
\usepackage{soul}
\setcitestyle{super}

\newtheorem{theorem}{Theorem}

\date{\today}
\begin{document}

\title{Playing nonlocal games with phases of quantum matter}

\author{Vir B. Bulchandani}
\affiliation{Princeton Center for Theoretical Science,  Princeton University, Princeton, New Jersey 08544, USA}
\affiliation{Department of Physics, Princeton University, Princeton, New Jersey 08544, USA}

\author{Fiona J. Burnell}
\affiliation{School of Physics and Astronomy, University of Minnesota, Minneapolis, Minnesota 55455, USA}

\author{S. L. Sondhi}
\affiliation{Rudolf Peierls Centre for Theoretical Physics, University of Oxford, Oxford OX1 3PU, United Kingdom}

\begin{abstract}

The parity game is an example of a nonlocal game: by sharing a Greenberger-Horne-Zeilinger (GHZ) state before playing this game, the players can win with a higher probability than is allowed by classical physics. The GHZ state of $N$ qubits is also the ground state of the ferromagnetic quantum Ising model on $N$ qubits in the limit of vanishingly weak quantum fluctuations. Motivated by this observation, we examine the probability that $N$ players who share the ground state of a generic quantum Ising model, which exhibits non-vanishing quantum fluctuations, still win the parity game using the protocol optimized for the GHZ state. Our main result is a modified parity game for which this protocol asymptotically exhibits quantum advantage in precisely the ferromagnetic phase of the quantum Ising model. We further prove that the ground state of the exactly soluble $d=1+1$ transverse-field Ising model can provide a quantum advantage for the parity game over an even wider region, which includes the entire ferromagnetic phase, the critical point and part of the paramagnetic phase. By contrast, we find examples of topological phases and symmetry-protected topological (SPT) phases of matter, namely the deconfined phase of the toric code Hamiltonian and the $\mathbb{Z}_2 \times \mathbb{Z}_2$ SPT phase in one dimension, that do not exhibit an analogous quantum advantage away from their fixed points.
\end{abstract}

\maketitle

\section{Introduction}
Bell's proof\cite{bell1964einstein} of his eponymous inequalities in 1964 was the first clear demonstration that the principles of quantum mechanics are incompatible with a description of reality that refers solely to local classical variables, an idea that is captured by the modern notion of contextuality\cite{Spekkens_2008,Abramsky_2011}. In subsequent decades, the essence of Bell's theorem has been distilled into ever simpler examples of quantum phenomena that are impossible to model classically. One particularly striking class of examples are nonlocal games, also known as ``quantum pseudo-telepathy games''\cite{Brassard2005}. These games are typically cooperative games, played by at least two players who are not allowed to communicate classically with one another.  Nonlocal games have the property that by sharing an entangled quantum state before playing the game, the players can win the game with a higher probability than the principles of spatial locality and classical physics allow.

The parity game, which was introduced by Brassard-Broadbent-Tapp\cite{brassard2005recasting} building on earlier work by Mermin\cite{mermin1990extreme}, is a relatively unusual example of a nonlocal game that is  ``scalable''\cite{Brassard2005}; for any $N \geq 2$, it is possible to define an $N$-player nonlocal parity game that can be won with certainty if the players are allowed to share an $N$-qubit Greenberger-Horne-Zeilinger (GHZ) state
\begin{equation}
\label{eq:GHZplus}
|\mathrm{GHZ}^{\pm}\rangle  = \frac{1}{\sqrt{2}}\left(|00 \ldots 0\rangle \pm |11\ldots 1\rangle\right)
\end{equation}
before playing the game.   This scalability property of the parity game resembles the scaling behaviour of extensive many-body quantum systems, and in fact the GHZ states in Eq. \eqref{eq:GHZplus} have a rather natural and well-known realization in condensed matter physics as the ground state doublet of a spin-$1/2$ quantum Ising (or $\mathbb{Z}_2$-symmetric) ferromagnet on $N$ qubits in the limit of vanishingly weak quantum fluctuations. 

However, condensed matter physics is usually concerned with \emph{phases} of matter that exhibit qualitatively similar properties over some non-zero range of model parameters. From this point of view, the GHZ states are non-generic, as they merely define a renormalization-group fixed point within the entire ferromagnetic phase of the quantum Ising model, which can exhibit strong quantum fluctuations away from this fixed point. We note that two other scalable nonlocal games, namely the multi-player triangle game\cite{Cluster,Bravyi} studied recently\cite{DanielMiyake}, and the toric code game proposed by the authors in a companion paper\cite{companion}, also admit perfect quantum strategies that make use of the fixed points of quantum phases of matter, though these strategies do not make use of conventional symmetry-breaking phases as arise in the quantum Ising model. Instead, these strategies involve the $\mathbb{Z}_2 \times \mathbb{Z}_2$ symmetry-protected topological (SPT) phase and the topological phase of the toric code respectively.

This raises the question of how far nonlocal games can be won with ground-state phases of matter \emph{away} from their fixed points. As discussed above, these phases might be either conventional symmetry-breaking phases or more exotic phases of matter. We note that this question was previously studied for the $\mathbb{Z}_2 \times \mathbb{Z}_2$-symmetric SPT phase\cite{DanielMiyake}. One fundamental difference between phases of matter and their fixed points is the presence of a non-zero, finite correlation length $0<\xi<\infty$ away from fixed points. The significance of this length scale for nonlocal games is that players who share entanglement with one another on length scales $\ell \ll \xi$ have greater access to multipartite entanglement within a given state than players who can only probe entanglement at larger length scales $\ell \gg \xi$, and might therefore be expected to perform better at nonlocal games. On the other hand, at length scales $\ell \ll \xi$, the underlying entanglement structure may depend sensitively on the microscopic properties of the model, and is not in general characteristic of the phase.

We believe that such ideas are potentially of broader interest, insofar as the advantage gained over classical physics in playing a specific multi-player game yields a measure of the ``quantumness'' of the many-body state in question, which probes its contextuality properties directly\cite{Abramsky2017} and to that extent is distinct\cite{EntUseless,ContextualityMagic} from popular measures of entanglement such as entanglement entropies. It is worth emphasizing that the problem of diagnosing and classifying the possible phases of many-body quantum matter through entanglement continues to inspire new theoretical developments\cite{kim2021chiral,siva2021universal}, despite the deep understanding of topological phases that has been achieved in recent decades\cite{StringNet,KitaevPreskill,DetStringNet,EntSpec,3d_class,SPTClass,Kane_2005,Moore_2007}. 

In order to formulate such questions mathematically, we define a \emph{quantum strategy} $\mathcal{S} = (|\psi\rangle,\mathcal{P})$ for an $N$-player nonlocal game $\mathcal{G}$ to consist of 
\begin{enumerate}
\item an $\mathcal{O}(N)$-qubit pure state $|\psi\rangle$, which is shared by all players before the game begins, with each player assigned to a specific qubit or set of qubits.
\item a \emph{protocol} $\mathcal{P}$, which is a set of operations that each player, or team of players\cite{companion}, applies to their qubits, and may consist of any sequence of quantum gates and measurements on those qubits.
\end{enumerate}
(The restriction to qubits is not binding\cite{Brassard2005}; see e.g. Appendix \ref{app:Boyer}.) We can then address the question of whether a given phase of matter yields quantum advantage for $\mathcal{G}$ away from fixed points as follows. Let $|\psi^*\rangle$ be the ground state of an $\mathcal{O}(N)$-body Hamiltonian representing the renormalization-group fixed point  of some quantum phase of matter, and suppose that the protocol $\mathcal{P}^*$ applied to the state $|\psi^*\rangle$ wins the nonlocal game $\mathcal{G}$ with certainty.  Thus the ``fixed-point quantum strategy'' $\mathcal{S}^* = (|\psi^*\rangle,\mathcal{P}^*)$ is a perfect quantum strategy\cite{brassard2005recasting} for $\mathcal{G}$.  Now consider a state $|\psi\rangle \neq |\psi^*\rangle$ that is the ground state of another Hamiltonian, but in the same phase of matter as $|\psi^*\rangle$ and with the same number of qubits. A natural way to explore whether $|\psi\rangle$ provides a quantum advantage for winning $\mathcal{G}$ away from the fixed point is by applying the protocol $\mathcal{P}^*$ to the state $|\psi\rangle$, rather than the state $|\psi^*\rangle$. Thus one studies the success rate of the quantum strategy $\mathcal{S} = (|\psi\rangle, \mathcal{P}^*)$. Generically, this strategy will be imperfect,  but may nevertheless provide a quantum advantage over the best possible classical strategy.  This is the approach that we shall pursue below (and was implicit in previous work\cite{DanielMiyake}).

Our goal in studying such imperfect quantum strategies will be two-fold. One strand of motivation comes from the theory of nonlocal games: for all such games studied in this paper, the best quantum strategy wins with probability $p^*_{\mathrm{qu}}=1$, while the best classical strategy wins with some probability $p^*_{\mathrm{cl}} < 1$. One might wonder whether it is possible to construct quantum strategies $\mathcal{S}$ whose probability of winning $p_{\mathrm{qu}}$ interpolates continuously between the optimal quantum value and the optimal classical value. Our analysis below yields an explicit solution to this problem, in the form of quantum strategies $\mathcal{S} = (|\psi\rangle,\mathcal{P}^*)$ where $|\psi\rangle$ is allowed to vary smoothly within the appropriate phase of matter. We find more generally that for the protocols studied in this work and for finite numbers of qudits, the quantum probability of winning $p_{\mathrm{qu}}(|\psi\rangle)$ is a continuous function of the state $|\psi\rangle$ (see e. g. Theorems \ref{thm:thm} and \ref{thm:thm2}), so that it is possible to construct quantum strategies that exhibit an arbitrarily small quantum advantage compared to $p_{\mathrm{cl}}^*$. An important question for future work is to understand how far the numerical value of $p_{\mathrm{qu}}(|\psi\rangle)$ can be related to existing measures of the quantumness of the state $|\psi\rangle$, such as the ``contextual fraction''\cite{Abramsky2017}, but we will not pursue this line of inquiry below.

Our second goal, which is more natural from a condensed matter perspective, will be to understand how far a quantum advantage for $\mathcal{S}=(|\psi\rangle,\mathcal{P}^*)$ constrains the state $|\psi\rangle$ to be in a given phase of matter. (Note that this is related to the question of whether nonlocality properties yield a useful probe for diagnosing quantum phase transitions\cite{BellNonloc}.) In exploring this question, we will primarily be concerned with the ``large-system limit'', in which the value of $N \gg 1$ can be taken to be arbitrarily large but finite. This should be contrasted with the infinite-system limit $N=\infty$, for which the mathematical assumption of ``cluster decomposition''\cite{weinberg1995quantum} of local observables is usually imposed in order to specify the representation of the infinite-dimensional algebra of local operators\cite{haag1964algebraic}. The cluster decomposition assumption forbids ground states that comprise macroscopically entangled superpositions of quantum states, and thus implies spontaneous symmetry breaking in the infinite-system limit. By working in the large-system limit instead, we can evade spontaneous symmetry breaking and define ground states that explicitly violate the cluster decomposition property for some (though not all) local observables.

Although this choice is in tension with the usual theoretical treatment of phases of matter, which are only strictly well-defined in the infinite-system limit, it is desirable for our purposes because it is specifically the ground states of \emph{finite} condensed matter systems that can exhibit macroscopic entanglement and therefore provide a useful resource for quantum games. Our choice is also consistent with the experimental realization of artificial condensed matter systems in the laboratory using cold atoms or arrays of qubits; the effective Hamiltonians for these systems are far from the infinite-system limit in which the assumption of cluster decomposition becomes mathematically useful, so that achieving macroscopic quantum superpositions becomes a difficult technical challenge rather than a theoretical impossibility\cite{Lukin,Song}.

The paper is structured as follows.  We first consider playing nonlocal games with conventional symmetry-breaking phases of matter, using the ferromagnetic phase of the one-dimensional quantum Ising model as our example. We describe the parity game and its perfect quantum strategy $\mathcal{S}_{\mathrm{BBT}}$ due to Brassard-Broadbent-Tapp (BBT)\cite{Brassard2005}. We then prove that the BBT protocol applied to the ground state of the one-dimensional transverse-field Ising model (TFIM) yields a quantum advantage for the parity game over a  non-zero range of transverse fields, which, for any number of players, includes and exceeds the entire ferromagnetic phase. One might wonder whether a nonlocal game can delineate the ferromagnetic phase of the quantum Ising model more clearly. We demonstrate that this can indeed be achieved by modifying the distribution of inputs to the parity game, and present a family of such modified parity games for which the BBT protocol exhibits quantum advantage (in an asymptotic sense) in precisely the ferromagnetic phase of the quantum Ising model. Extensions to states with $\mathbb{Z}_M$ symmetry, for example the ground states of clock models, are discussed in Appendix \ref{app:Boyer}.

We next turn to topological phases of matter. We discuss the toric code game introduced in a companion paper\cite{companion}, and argue that in the deconfined phase arising when the ideal toric code Hamiltonian is perturbed by weak magnetic fields\cite{KITAEV20032,Tupitsyn_2010}, the fixed-point protocol fails to yield a quantum advantage away from the ideal toric code fixed point. We show that this failure is a consequence of the perimeter-law scaling of the expectation values of the Wilson loop operators with which the toric code game is played\cite{companion}, suggesting a basic distinction between the robustness to perturbations of quantum strategies for nonlocal games that involve local operators, versus strategies that involve only nonlocal operators.

Finally, we revisit the $\mathbb{Z}_2 \times \mathbb{Z}_2$ SPT phase in one dimension and its relation to the triangle game, which was studied in previous work\cite{DanielMiyake}. We introduce a family of scalable nonlocal games associated with matchings of polygons that extend both the triangle game and the multiplayer triangle game and can be won with certainty using the fixed point of the phase. Nevertheless, we argue that these ``polygon games'', including the triangle game, are insufficient to uniquely determine the $\mathbb{Z}_2 \times \mathbb{Z}_2$ SPT phase or even its fixed point.

\section{The ferromagnetic phase of the quantum Ising model}
\label{sec:Ising}
\subsection{Playing the parity game with the Ising ground state}
\subsubsection{The parity game}
The parity game is played as follows\cite{brassard2005recasting}. There are $N \geq 3$ players and player $j$ is given a classical bit $a_j \in \{0,1\}$, with the promise that $\sum_{j=1}^N a_j$ is even. In order to win the game, the players must output bits $b_j \in \{0,1\}$ such that
\begin{equation}
\sum_{j=1}^N b_j \equiv \frac{\sum_{j=1}^N a_j}{2} \mod{2}.
\end{equation}
The players may not communicate classically with one another, but they are allowed to share an $N$-qubit quantum state $|\psi\rangle$ before playing the game. During the course of the game, the $j$th player is free to apply quantum gates and projective measurements to the $j$th qubit.

If the players do not take advantage of quantum physics, the optimal strategy available to them wins with a probability $p_{\mathrm{cl}}^*<1$ given by\cite{brassard2005recasting}
\begin{equation}
\label{eq:classwinprob}
p_{\mathrm{cl}}^* = \frac{1}{2} + \frac{1}{2^{\lceil N/2\rceil}}
\end{equation}
for inputs $\{a_j\}_{j=1}^N$ chosen randomly and uniformly from the set of $2^{N-1}$ bit strings fulfilling the promise. If the players instead share the quantum state $|\psi\rangle = |\mathrm{GHZ}^+\rangle$ before playing the game, then they can win the game with probability $p_{\mathrm{qu}} = 1$ by performing the following three operations, which we refer to collectively as the BBT protocol, $\mathcal{P}_{\mathrm{BBT}}$:
\begin{enumerate}
    \item Each player acts on their spin with the phase gate $\hat{Z}_j^{a_j/2} = \begin{pmatrix} 1 & 0 \\ 0 & i^{a_j} \end{pmatrix}$.
    \item Each player rotates to the $\hat{X}$ or Hadamard basis by applying the gate
    \begin{equation}
    \hat{U} = \frac{1}{\sqrt{2}} \begin{pmatrix} 1 & 1 \\ 1 & -1 \end{pmatrix}.
    \end{equation}
    \item Each player measures their qubit in the $\hat{Z}$ or computational basis and returns the outcome of their measurement $b_j \in \{0,1\}$.
\end{enumerate}
Let us briefly derive the above result. After Step 1, the state $|\psi\rangle$ is mapped to
\begin{equation}
|\psi'\rangle = \begin{cases} \frac{1}{\sqrt{2}}\left(| 00 \ldots 0\rangle + | 11 \ldots 1 \rangle\right) &  \frac{\sum_{j=1}^N a_j}{2}\,\, \mathrm{even} \\
\frac{1}{\sqrt{2}}\left(| 00 \ldots 0\rangle - | 11 \ldots 1\rangle\right) &  \frac{\sum_{j=1}^N a_j}{2} \,\, \mathrm{odd}
\end{cases}
\end{equation}
After Step 2, this becomes
\begin{equation}
|\psi''\rangle \propto \begin{cases} \sum_{\{\vec{b} : \sum_{j=1}^N b_j \, \mathrm{even}\}} |b_1 b_2 \ldots b_N \rangle & \frac{\sum_{j=1}^N a_j}{2}\,\, \mathrm{even} \\
\sum_{\{\vec{b} : \sum_{j=1}^N b_j \, \mathrm{odd}\}} |b_1 b_2 \, \ldots b_N \rangle & \frac{\sum_{j=1}^N a_j}{2}\,\, \mathrm{odd}
\end{cases}
\end{equation}
which is an equal weight superposition of states with the required parity. It follows that upon performing Step 3, a shared state $|b_1\ldots b_N\rangle$ is obtained with $\sum_{j=1}^N b_j \equiv \frac{\sum_{j=1}^N a_j}{2} \mod{2}$ as desired.
Thus $N$ players who share the state $|\mathrm{GHZ}^+\rangle$ can win the parity game for any allowed input $\{a_j\}_{j=1}^N$.

\subsubsection{Playing with Ising ground states}
As noted in the introduction, the GHZ state is the ground state of the quantum Ising model in the limit of vanishingly weak quantum fluctuations. We would now like to consider what happens to the efficacy of the BBT protocol when quantum fluctuations are no longer weak. Specifically, we consider the probability of winning the parity game with the ground state of the TFIM with an additional longitudinal field---sometimes called the tilted field Ising model---on $N$ qubits, namely
\begin{equation}
\label{eq:LTFIM}
\hat{H} = -J \sum_{j=1}^N \hat{Z}_j \hat{Z}_{j+1} - \Gamma \sum_{j=1}^N \hat{X}_j - h \sum_{j=1}^N \hat{Z}_j.
\end{equation}
We set $J,\, \Gamma > 0$, assume periodic boundary conditions with $\hat{Z}_{N+1} \equiv \hat{Z}_1$, and treat the cases $h \neq 0$ and $h=0$ separately. As is well known, the ground state of this model $|\psi_0\rangle$ coincides with the $|\mathrm{GHZ}^+\rangle$ state in the limit $h=0,\, \Gamma \to 0^+$. The question to be addressed is whether the ground state $|\psi_0\rangle$ can be used to win the parity game \emph{away} from this limit using the BBT protocol. (We will comment later on possible alternative quantum strategies based on different protocols $\mathcal{P} \neq \mathcal{P}_{\mathrm{BBT}}$.)

In seeking to answer this question, we have recourse to the following general result (which is proved in Appendix \ref{app:Theorem1}):

\begin{theorem}
\label{thm:thm}
The quantum strategy $\mathcal{S} = (|\psi\rangle, \mathcal{P}_{\mathrm{BBT}})$ wins the parity game with probability
\begin{equation}
\label{eq:exactquantum}
p_{\mathrm{qu}}(|\psi\rangle) = \frac{1}{2}\left(1+ |\langle \psi | \mathrm{GHZ}^+\rangle |^2 -|\langle \psi | \mathrm{GHZ}^-\rangle |^2\right),
\end{equation}
that depends solely on the fidelity of $|\psi\rangle$ to the $N$-qubit GHZ states.
\end{theorem}
Whenever this probability exceeds the optimal classical probability of winning, Eq. \eqref{eq:classwinprob}, we shall say that $|\psi\rangle$ provides a ``quantum advantage'' over the optimal classical strategy.

The remainder of this subsection is structured as follows. We first use mean-field reasoning to argue that quantum advantage of the Ising ground state  $|\psi_0\rangle$ is lost for any non-zero $\mathbb{Z}_2$-symmetry-breaking longitudinal field $h \neq 0$. This leads us to consider the pure transverse-field Ising model with $h=0$,
\begin{equation}
\label{eq:TFIM}
\hat{H}_{\mathrm{TFIM}} = -J \sum_{j=1}^N \hat{Z}_j \hat{Z}_{j+1} - \Gamma \sum_{j=1}^N \hat{X}_j,
\end{equation}
for which a mean-field analysis of the ferromagnetic phase $\Gamma \leq J$ suggests persistence of quantum advantage up to a non-zero value $0<\Gamma_* < J$ as $N \to \infty$ in the large-system limit. Finally, we use exact solvability of the TFIM Eq. \eqref{eq:TFIM} to both confirm the persistence of quantum advantage throughout the ferromagnetic phase $\Gamma \leq J$, and determine the precise threshold at which the TFIM ground state ceases to provide a quantum advantage for winning the parity game in the large-system limit, which turns out to lie outside the ferromagnetic phase and is found to occur at a transverse field strength $\Gamma_* \approx 1.506 J$. (We note that outside the ferromagnetic phase $\Gamma > J$, there is a unique ground state as $N \to \infty$, and therefore no need to distinguish carefully between the large-system and infinite-system limits.)

\subsubsection{Mean-field predictions}
\label{sec:MFP}
The GHZ states are only exact ground states of the generic Ising Hamiltonian \eqref{eq:LTFIM} at the point $h=\Gamma=0$. At this point the model is ferromagnetic, so to analyze the ground states in its vicinity, we perform a mean-field analysis in a ferromagnetic background $\langle \hat{Z}_i \rangle = m$. In Eq. \eqref{eq:LTFIM}, this yields the mean-field Hamiltonian
\begin{equation}
\hat{H}^{\mathrm{MF}} = \sum_{j=1}^N -(mJ +h)\hat{Z}_j -\Gamma \hat{X}_j = \sum_{j=1}^N \hat{h}_j^{\mathrm{MF}},
\end{equation}
where the onsite mean-field Hamiltonian is given by
\begin{equation}
\label{eq:onsite}
\hat{h}_j^{\mathrm{MF}} = \begin{pmatrix} -(mJ+h) & -\Gamma \\ -\Gamma & mJ+h \end{pmatrix}.
\end{equation}
For $h \neq 0$, the onsite Hamiltonian has a non-degenerate ground state with energy $\epsilon = -\sqrt{(mJ+h)^2 + \Gamma^2}$ and eigenvector
\begin{equation}
|v\rangle = \begin{pmatrix} \cos{\frac{\theta}{2}} \\ \sin{\frac{\theta}{2}} \end{pmatrix}, \quad \theta = \tan^{-1}{\frac{\Gamma}{mJ+h}}.
\end{equation}
The magnetization $m$ is determined self-consistently by the relation $\langle v| \hat{Z} |v\rangle = \cos{\theta} = m$. The mean-field ground state is given by
\begin{equation}
\label{eq:SBMF}
|\psi_{\mathrm{MF}} \rangle = \otimes_{i=1}^N |v\rangle,
\end{equation}
whence it follows by Eq. \eqref{eq:exactquantum} that the probability of winning the parity game with the BBT protocol applied to this state is
\begin{equation}
p_{\mathrm{qu}} = \frac{1}{2} + \cos^N{\frac{\theta}{2}}\sin^N{\frac{\theta}{2}}.
\end{equation}
Since
\begin{equation}
\cos^N{\frac{\theta}{2}}\sin^N{\frac{\theta}{2}} = \frac{1}{2^N} \sin^N{\theta} < \frac{1}{2^{\lceil N/2 \rceil}},
\end{equation}
we have shown that for the mean-field ground state with $h \neq 0$, quantum advantage is always lost, i.e.
\begin{equation}
p_{\mathrm{qu}} < p_{\mathrm{cl}}^*.
\end{equation}

Let us therefore set $h=0$ and turn to a mean-field analysis of the TFIM Hamiltonian, Eq. \eqref{eq:TFIM}. In this case, the onsite mean-field Hamiltonian Eq. \eqref{eq:onsite} exhibits a degeneracy between magnetizations $\pm |m|$. This yields two degenerate ground states
\begin{equation}
|v_+\rangle = \begin{pmatrix} \cos{\frac{\theta}{2}} \\ \sin{\frac{\theta}{2}}\end{pmatrix}, \, |v_-\rangle= 
\begin{pmatrix} \sin{\frac{\theta}{2}} \\ \cos{\frac{\theta}{2}}\end{pmatrix},\, \tan{\theta} = \frac{\Gamma}{|m|J}
\end{equation}
corresponding to a given value of $|m|$. For both cases, the self-consistent magnetization is determined by $\cos{\theta} = |m|$, i.e.
\begin{equation}
\label{eq:selfconsmag}
|m| = \sqrt{1-(\Gamma/J)^2}.
\end{equation}
Thus a translation-invariant, even-parity mean-field ground state for the TFIM is given by
\begin{equation}
|\psi_{\mathrm{MF}}\rangle = \mathcal{N} (\otimes_{i=1}^N |v_+\rangle+\otimes_{i=1}^N |v_-\rangle),
\end{equation}
where the normalization constant satisfies $2(1+\sin^N{\theta})\mathcal{N}^2=1$. This implies a probability of winning the parity game
\begin{equation}
\label{eq:TFIMMF}
p_{\mathrm{qu}} = \frac{1}{2} + \frac{1}{2(1+\sin^{N}{\theta})} \left(\cos^{N}{\frac{\theta}{2}} +\sin^{N}{\frac{\theta}{2}}\right)^2.
\end{equation}
Letting $g = \Gamma/J$ and using the self-consistent value of $\theta$, this can be written as
\begin{align}
\nonumber
&p_{\mathrm{qu}} = 
\frac{1}{2}+ \\
&\frac{1}{1 + g^N} \left[\left(\frac{g}{2}\right)^N + \frac{(1 + \sqrt{1-g^2})^N + (1 - \sqrt{1-g^2})^N}{2^{N+1}}\right].
\end{align}

\subsubsection{A battle of exponentials}
\label{sec:boe}
Notice that as $N\to \infty$, the probability of winning the parity game with the BBT protocol applied to the mean-field ground state of the TFIM exhibits a jump discontinuity at $g=0$, with
\begin{equation}
\label{eq:jump}
p_{\mathrm{qu}} \to
\begin{cases} 1 & g=0 \\  \frac{1}{2} & g >0 \end{cases}, \quad N \to \infty.
\end{equation}
It thus appears that in the large-system limit, any advantage over the optimal classical strategy is lost as $N \to \infty$, for any $g>0$. However, this order of limits is too crude for the problem at hand, since as $N \to \infty$, it is possible for the quantum strategy to outperform the best classical strategy for any finite $N$, even at non-zero $g$. 

In particular, quantum advantage can persist for $g>0$ as $N \to \infty$ if the exponentially small correction to random guessing for the quantum strategy,
$p_{\mathrm{qu}}-1/2$, decays more slowly in $N$ than the correction $p_{\mathrm{cl}}^*-1/2 = 1/2^{\lceil N/2\rceil}$ for the best classical strategy, i.e. if the quantum strategy beats the optimal classical strategy in this ``battle of exponentials''. We now show that the BBT protocol applied to the mean-field TFIM ground state wins the battle of exponentials in most of the ferromagnetic phase.

Focusing on the leading correction to random guessing in Eq. \eqref{eq:TFIMMF}, we find that
\begin{equation}
p_{\mathrm{qu}} - \frac{1}{2} \sim  \frac{1}{2}\left(\frac{1+\sqrt{1-g^2}}{2}\right)^N, \quad N \to \infty,
\end{equation}
with errors that are exponentially smaller in $N$ than this leading term. Comparison with the classical result as $N \to \infty$ implies a loss of quantum advantage when $g$ reaches the threshold 
\begin{equation}
g \geq g^{\mathrm{MF}}_* = \sqrt{2}{\sqrt{\sqrt{2}-1}} \approx 0.910,
\end{equation}
which by Eq. \eqref{eq:selfconsmag} is equivalent to a threshold for the local magnetization
\begin{equation}
\label{eq:MFmagcondition}
m \leq m^{\mathrm{MF}}_* = \sqrt{2}-1.
\end{equation}
This prediction of the mean-field analysis raises the intriguing possibility that the ground state of the TFIM provides a quantum advantage for winning the parity game over the entire ferromagnetic phase. In fact, this prediction is confirmed by the exact solution to the model, as we demonstrate below. Surprisingly, quantum advantage persists even beyond the critical point, and extends some distance into the paramagnetic phase.

\subsubsection{Exact results for the TFIM}

\begin{figure}[t]
    \centering
    \includegraphics[width=0.95\linewidth]{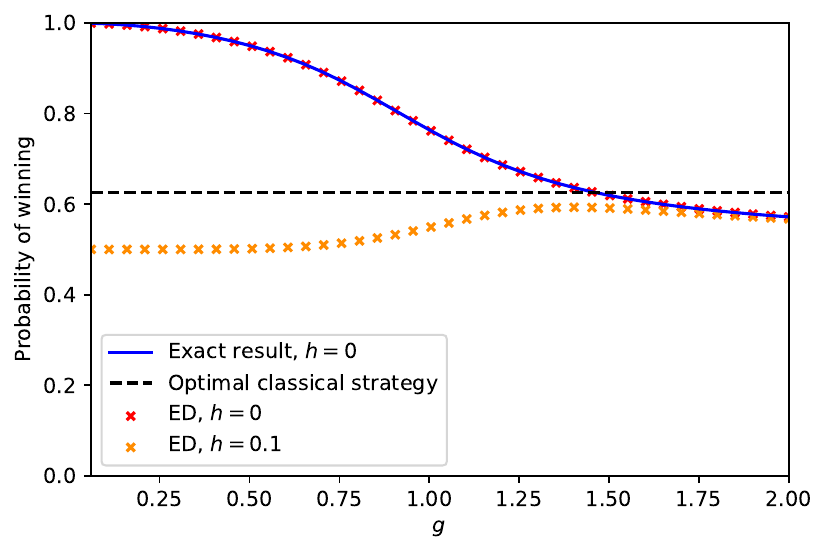}
    \caption{Exact diagonalization of the Ising model Eq. \eqref{eq:LTFIM} on $N=6$ qubits with $J=1$ fixed and $\Gamma$ and $h$ varied. The probability of winning the parity game is obtained numerically from Eq. \eqref{eq:exactquantum}. The exact result Eq. \eqref{eq:exactTFIM} matches the numerics for $h=0$ as expected, and quantum advantage is sustained up to a transverse field strength $g_* \approx 1.5$. For a small longitudinal field $h=0.1$, quantum advantage is lost, which is consistent with the mean-field analysis of Section \ref{sec:MFP}.}
    \label{Fig1}
\end{figure}

We now report an analytical result for the probability of winning the parity game with the BBT protocol applied to the ground state of the transverse field Ising model Eq. \eqref{eq:TFIM}. Our starting point is the exact ground-state wavefunction of the model, written in terms of pairs of Jordan-Wigner fermions (we refer to Ref. \onlinecite{Damski_2013} for details). For all $N$ and $\Gamma>0$, this is given by the even-parity ground state
\begin{equation}
|\psi_0 \rangle = \prod_{k>0} \left(\cos{\frac{\theta_k}{2}} - \sin{\frac{\theta_k}{2}}c_k^\dagger c_{-k}^\dagger \right) |0\rangle,
\end{equation}
where $c_k|0\rangle = 0$ is a vacuum of Jordan-Wigner fermions, the wavenumbers that are occupied in this state are given by
\begin{equation}
k = \begin{cases} \pm \frac{\pi}{N}, \pm \frac{3\pi}{N}, \ldots \pm \frac{(N-1)\pi}{N},  & N \mathrm{even} \\ 
\pm \frac{\pi}{N}, \pm \frac{3\pi}{N}, \ldots \pm \frac{(N-2)\pi}{N}, & N \, \mathrm{odd}\end{cases},
\end{equation}
and the single-particle energies $\epsilon_k = \sqrt{\Gamma^2+J^2-2\Gamma J \cos{k}}$ determine the Bogoliubov angles via
\begin{equation}
\label{eq:bog}
\sin{\theta_k} = \frac{J \sin{k}}{\epsilon_k}, \quad \cos{\theta_k} = \frac{\Gamma-J \cos{k}}{\epsilon_k}.
\end{equation}
The key observation for analytically calculating the probability of winning the parity game with this state is that the fidelity appearing in Eq. \eqref{eq:exactquantum} can be expressed as an overlap of pairing wavefunctions. In more detail, degenerate perturbation theory implies that $\lim_{\Gamma\to 0^+} |\psi_0\rangle = |\mathrm{GHZ}^+\rangle$, while $|\mathrm{GHZ}^{\pm}\rangle$ states have parity $\hat{P} = \pm 1$, so that Eq. \eqref{eq:exactquantum} reduces to
\begin{equation}
p_{\mathrm{qu}} = \frac{1}{2}\left(1+|\langle \psi_0(\Gamma)|\psi_0(\Gamma=0^+)\rangle|^2\right).
\end{equation}
This fidelity can be written down explicitly in terms of the Bogoliubov angles as
\begin{equation}
|\langle \psi_0(\Gamma)|\psi_0(\Gamma=0^+)\rangle|^2 = \prod_{k>0} \cos^2{\left(\frac{\theta_k(\Gamma) - \theta_k(\Gamma=0^+)}{2}\right)}.
\end{equation}
Some elementary trigonometric manipulations and Eq. \eqref{eq:bog} then imply that
\begin{equation}
\label{eq:exactTFIM}
p_{\mathrm{qu}} = \frac{1}{2}+\frac{1}{2^{\lfloor N/2\rfloor}}\prod_{k>0}\left(1+ \frac{1-g \cos{k}}{\sqrt{1+g^2-2 g \cos{k}}}\right),
\end{equation}
which should be compared to the optimal classical probability of winning Eq. \eqref{eq:classwinprob}. As for the mean-field ground state discussed above, the possibility of quantum advantage is determined by a competition between corrections to random guessing that are both exponentially small in $N$. The threshold value $g_*$ at which the Ising ground state loses this battle of exponentials satisfies the transcendental equation
\begin{equation}
\prod_{k>0}\left(1+ \frac{1-g_* \cos{k}}{\sqrt{1+g_*^2-2g_* \cos{k}}}\right) = \begin{cases} 1 & N \, \mathrm{even} \\ \frac{1}{2} & N \, \mathrm{odd}\end{cases}
\end{equation}
This result is exact for any finite $N$. In the large-system limit as $N\to \infty$, the left-hand side tends to an infinite product; taking logarithms before the large-$N$ limit yields the following equation for $g_*$:
\begin{equation}
\label{eq:numerically}
\int_{0}^\pi dk \, \log{\left(1 + \frac{1-g_* \cos{k}}{\sqrt{1+g_*^2-2g_* \cos{k}}}\right)} = 0.
\end{equation}
We note that for $g_* \leq 1$, the integrand is strictly positive almost everywhere so this equation has no solution in the ferromagnetic phase; this proves that the TFIM ground state yields an advantage over the best classical strategy as $N \to \infty$ in the entire ferromagnetic phase. In fact, a degree of quantum advantage persists \emph{beyond} the ferromagnetic phase; by solving Eq. \eqref{eq:numerically} numerically, we find that the limiting value of $g_*$ is given by 
\begin{equation}
\label{eq:exactg}
g_* \to 1.506..., \quad N \to \infty.
\end{equation}
We deduce that ground states of the TFIM can be used to accomplish tasks that are classically impossible, even in the paramagnetic phase of this model.

At first sight, this result is surprising, but it can be motivated in a more elementary fashion by perturbation theory about the strongly paramagnetic limit, $J=0, \Gamma>0$, where the ground state is the fully $\hat{X}$-polarized state 
\begin{equation}
|\mathrm{X}\rangle = \frac{1}{2^{N/2}} \otimes_{i=1}^N (|0\rangle + |1\rangle).
\end{equation}
In this limit, the probability of winning the parity game with the quantum strategy $\mathcal{S} = (|\mathrm{X}\rangle, \mathcal{P}_{\mathrm{BBT}})$ has the following intuitive interpretation: the BBT protocol applied to $|\mathrm{X}\rangle$ wins the parity game just as often as random guessing for input bit strings $\vec{a} \neq (0,0,\ldots,0)$ with at least one non-zero element, but wins with certainty if the input bits are identically zero, $\vec{a} = (0,0,\ldots,0)$. Averaging uniformly over input bit strings consistent with the promise then yields
\begin{equation}
p_{\mathrm{qu}}(|\mathrm{X}\rangle) = \frac{1}{2^{N-1}}\left(1+(2^{N-1}-1)\cdot \frac{1}{2}\right) = \frac{1}{2} + \frac{1}{2^N},
\end{equation}
which matches a direct calculation of Eq. \eqref{eq:exactquantum} for this state. 

Let us now consider the effect of introducing a weak ferromagnetic coupling $J>0$ and applying the BBT protocol to the resulting ground state $|\psi_0\rangle$. By the exact calculation above, we know that this quantum strategy provides a quantum advantage for sufficiently large $J$; we will now show this perturbatively in $J$. At leading order in $J/\Gamma = 1/g$, the perturbed ground state is given by
\begin{equation}
|\psi_{0}\rangle = |\mathrm{X}\rangle + \frac{J}{4\Gamma} \sum_{i=1}^N \hat{Z}_{i} \hat{Z}_{i+1} |\mathrm{X}\rangle + \mathcal{O}(J^2/\Gamma^2).
\end{equation}
In Eq. \eqref{eq:exactquantum}, this implies a probability of winning
\begin{equation}
p_{\mathrm{qu}} = \frac{1}{2} + \frac{1}{2^N}\left(1+\frac{N}{2g}\right) + \mathcal{O}(g^{-2}) \approx \frac{1}{2} + \frac{e^{N/2g}}{2^N}.
\end{equation}
Comparison with Eq. \eqref{eq:classwinprob} reveals that this perturbative estimate wins the battle of exponentials against the best classical strategy when
\begin{equation}
e^{N/2g} \sim 2^{N/2}
\end{equation}
which yields an estimate for the loss of quantum advantage when
\begin{equation}
g \geq g^{\mathrm{PT}}_* = 1/\log{2} \approx 1.443
\end{equation}
in the large-system limit. This estimate for the threshold $g_*$ lies firmly within the paramagnetic phase and (perhaps fortuitously) is accurate to within $5\%$ of the exact result, Eq. \eqref{eq:exactg}. This raises the possibility that in higher dimensions, where the TFIM is no longer exactly solvable, perturbation theory might be a useful tool for estimating $g_*$.

We conclude our theoretical analysis with a numerical simulation that corroborates it. In order for the battle of exponentials to be visible on a plot, it is preferable to model a small number of players. We therefore consider $N=6$ players who play the parity game with the BBT protocol applied to the ground state of a generic Ising model Eq. \eqref{eq:LTFIM} on 6 qubits, that is obtained by exact diagonalization. For a vanishing longitudinal field $h=0$ we find quantum advantage up to a threshold $g \approx 1.5$ that already approximates the large-system limit of $g_*$. For a small but non-zero longitudinal field $h=0.1$, quantum advantage is lost for all $\Gamma>0$, as expected from the mean-field analysis of Section \ref{sec:MFP}. See Fig. \ref{Fig1}.

\subsection{Nonlocal games for the ferromagnetic phase}
\label{sec:Pbitparity}

The above results lead to the \emph{a priori} unexpected conclusion that for any finite $N$, the TFIM ground state can provide a quantum advantage for the parity game firmly outside the ferromagnetic phase of the model. At the same time, this property is somewhat undesirable from the viewpoint of using quantum games to identify phases of matter. Meanwhile, a practical shortcoming of the above approach is that away from the point $\Gamma = 0$, the quantum advantage we obtain for the parity game strictly vanishes in the large-system limit, even in the ferromagnetic phase. This raises the question of whether one can use nonlocal games to more clearly discern the ferromagnetic phase.

In this subsection, we present a family of nonlocal games for which the BBT protocol applied to the Ising ground state yields a quantum advantage that is non-zero in the large-system limit in most of the ferromagnetic phase of the quantum Ising model. We further show that a family of such games can be constructed which lose quantum advantage at precisely the Ising quantum critical point, in an asymptotic sense to be explained below.

\subsubsection{The $N$-player $P$-bit parity game}

\begin{figure}[t]
    \centering
    \includegraphics[width=0.99\linewidth]{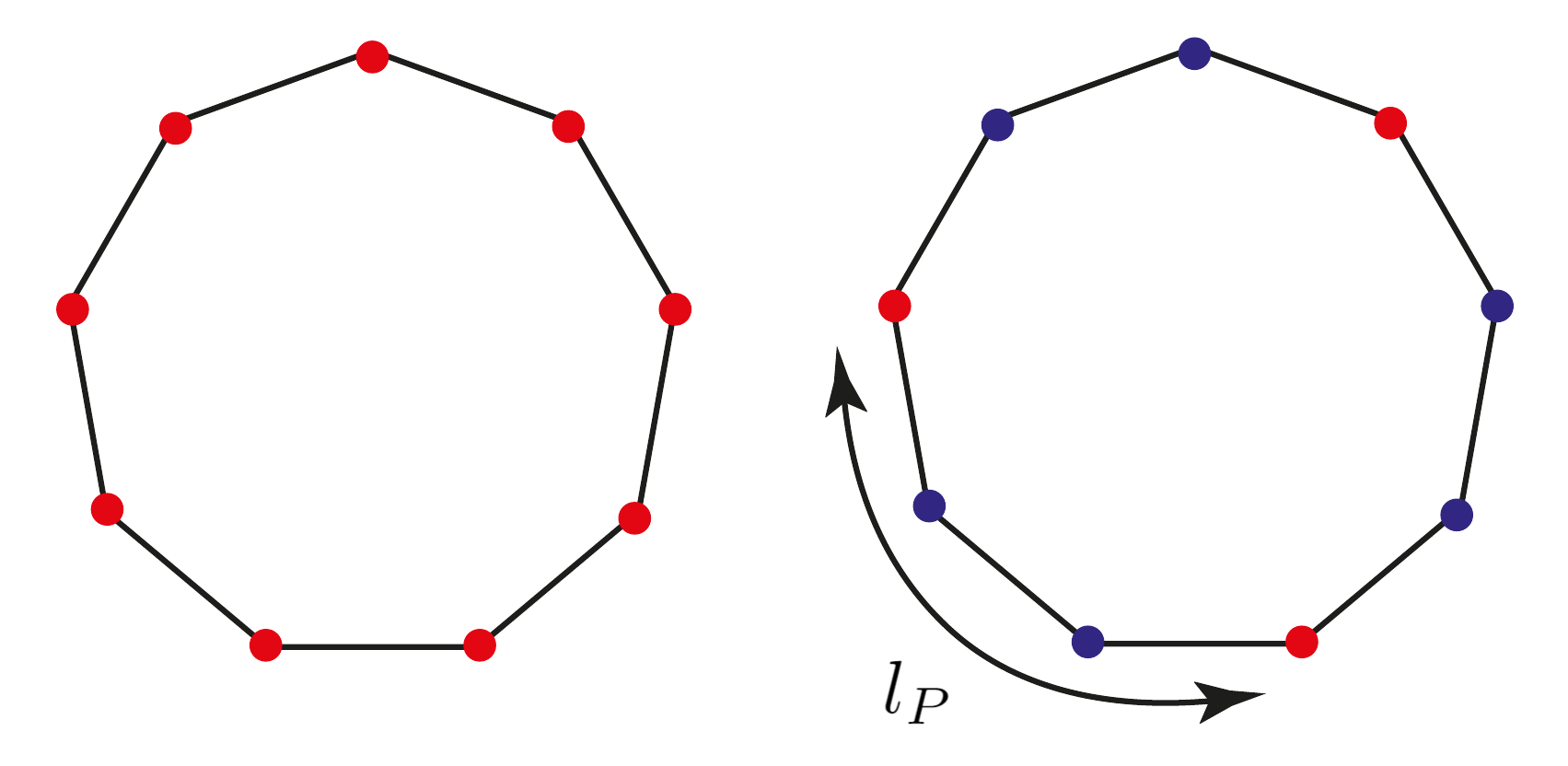}
    \caption{Schematic comparison between the usual parity game (\textit{left}) and the $P$-bit parity game (\textit{right}). Each player is depicted as a vertex on a polygon. In the usual parity game, every player can receive a bit $a_i \in \{0,1\}$. In the $P$-bit parity game, only marked players (depicted as red vertices) separated by a characteristic distance $l_P$ can receive non-zero input bits $a_i \in \{0,1\}$. All other players (blue vertices) receive an input bit $a_i=0$. In the figure we have set $N=9$ and $P=3$.}
    \label{Fig2}
\end{figure}

Consider a parity game with $N$ players, but now with a constraint on the input bit-strings, so that only $3 \leq P < N$ players can receive non-zero input bits, i.e. $P$ ``marked'' players $i_1,i_2,\ldots,i_P$ receive a bit $a_{i_j}\in\{0,1\}$ uniformly at random, subject to the promise that $\sum_{j=1}^P a_{i_j}$ is even, and the remaining $N-P$ players receive bits $a_i = 0$. This modification of the parity game is depicted in Fig. \ref{Fig2}. We will assume that before the game begins, each player knows whether they are allowed to receive a non-zero bit\footnote{Notice that if the players do not know before playing the game whether they will be marked, this does not change the best possible classical probability of winning, as one can always consider the optimal strategy on the marked bits with all other players returning zero.}. For $P=3$, this game bears a similar relation to the original parity game\cite{brassard2005recasting} as the multi-player triangle game bears to the triangle game\cite{Cluster,Bravyi,DanielMiyake}.

The Brassard-Broadbent-Tapp protocol for this constrained parity game, applied to an arbitrary state
\begin{equation}
|\psi\rangle = \sum_{\vec{\sigma}\in\{0,1\}^N} c_{\vec{\sigma}} |\vec{\sigma}\rangle,
\end{equation}
wins with probability
\begin{equation}
\label{eq:exactP}
p_{\mathrm{qu}}(|\psi\rangle) = \frac{1}{2}\left(1+ \sum_{\{\vec{\sigma}\in\{0,1\}^N:\sigma_{i_j} = 0\}} c_{\vec{\sigma}}c^*_{\vec{1}-\vec{\sigma}} + c_{\vec{1}-\vec{\sigma}}c^*_{\vec{\sigma}} \right),
\end{equation}
by arguments analogous to the proof of Theorem \ref{thm:thm}. (Here $\{i_j\}_{j=1}^P$ label the qubits of the $P$ marked players.) This can be written in terms of a GHZ stabilizer operator $\hat{B}$,
\begin{equation}
p_{\mathrm{qu}}(|\psi\rangle) = \frac{1}{2}\left(1+ \langle \psi| \hat{B}|\psi\rangle \right)
\end{equation}
with
\begin{equation}
\hat{B} = \left(\prod_{j=1}^P \left(\frac{1+\hat{Z}_{i_j}}{2}\right) + \prod_{j=1}^P \left(\frac{1-\hat{Z}_{i_j}}{2}\right)\right) \prod_{i=1}^N \hat{X}_i.
\end{equation}
We note that this result can be derived directly by observing that the BBT protocol amounts to measuring Mermin's stabilizers for the GHZ state\cite{mermin1990extreme}, then using the fact that these stabilizers define ``dichotomic observables''\cite{DanielMiyake}.

\subsubsection{$3$-bit parity games with TFIM ground states}
For the $3$-bit parity game played with an $N$-qubit TFIM ground state $|\psi_0\rangle$, the above result reduces to
\begin{equation}
\label{eq:exact3}
p_{\mathrm{qu}}(|\psi_0\rangle) = \frac{1}{8}\left(5+ \langle \hat{Z}_{i_1}\hat{Z}_{i_2}\rangle_0 + \langle \hat{Z}_{i_1}\hat{Z}_{i_3}\rangle_0 +\langle \hat{Z}_{i_2}\hat{Z}_{i_3} \rangle_0\right),
\end{equation}
where we introduced the notation $\langle \hat{O}\rangle_0 =  \langle \psi_0 | \hat{O}|\psi_0\rangle$ for ground-state expectation values of operators $\hat{O}$. At large length scales, the values of the ground-state correlation functions for the TFIM are well known\cite{pfeuty1970one}. To be precise, if the marked players $i_1,\,i_2,\,i_3$ are equally spaced on a ring of length $N$, then
\begin{equation}
p_{\mathrm{qu}}(|\psi_0\rangle)\to \begin{cases} \frac{5}{8} & g > 1\\ \frac{5}{8}+\frac{3}{8}(1-g^2)^{1/4} & g\leq 1\end{cases}
\end{equation}
in the large-system limit $N \to \infty$. Since we assumed that each player knows whether they are allowed to receive a non-zero bit before the game begins, the optimal classical strategy reduces to the optimal classical strategy on three sites, with all other players returning zero, which has classical winning probability
\begin{equation}
p_{\mathrm{cl}}^* = 3/4.
\end{equation}
This reveals that in the large-system limit, quantum advantage is lost when
\begin{equation}
g \geq g_*=\sqrt{\frac{80}{81}} \approx 0.994.
\end{equation}
Thus the BBT protocol applied to TFIM ground state provides a quantum advantage for the 3-bit parity game in $99.4 \%$ of the ferromagnetic phase.

Before developing this observation further, let us briefly consider playing the 3-bit parity game with the ground state of the TFIM in a longitudinal symmetry-breaking field, Eq. \eqref{eq:LTFIM}. In this case, the formula Eq. \eqref{eq:exact3} must be modified to explicitly include the parity operator $\hat{P} =\prod_{j=1}^N \hat{X}_j$, to yield
\begin{equation}
\label{eq:exact3long}
p_{\mathrm{qu}}(|\psi_0\rangle) = \frac{1}{8}\left(4+ \langle \hat{P}\rangle_0 + \sum_{j<k} \langle \hat{Z}_{i_j}\hat{Z}_{i_k}\hat{P}\rangle_0 \right).
\end{equation}
As in the previous subsection, we can estimate this quantity using mean-field theory. For the mean-field symmetry-broken ground state Eq. \eqref{eq:SBMF}, we have
\begin{equation}
\langle \psi_{\mathrm{MF}} | \hat{Z}_{i_j} \hat{Z}_{i_k} \hat{P} | \psi_{\mathrm{MF}} \rangle = 0, \quad j\neq k,
\end{equation}
and
\begin{equation}
\quad \langle \psi_{\mathrm{MF}} |\hat{P} | \psi_{\mathrm{MF}} \rangle = \sin^{N}\theta.
\end{equation}
It follows by Eq. \eqref{eq:exact3long} that
\begin{equation}
p_{\mathrm{qu}}(|\psi_{\mathrm{MF}}\rangle) = \frac{1}{2}+\frac{1}{8}\sin^{N}{\theta} \to \frac{1}{2}, \quad N \to \infty.
\end{equation}
Thus a mean-field analysis predicts no quantum advantage in the large-system limit. We expect the same conclusion to hold for all other $P$-bit parity games with $3 < P < N$.

\subsubsection{$P$-bit parity games with TFIM ground states}

One merit of the $3$-bit parity game for our purposes, compared to the standard $N$-bit parity game, is that the level of quantum advantage provided by the BBT protocol, as measured by the quantity $\Delta p = p_{\mathrm{qu}}(|\psi_0\rangle) - p_{\mathrm{cl}}^*$, remains finite and non-zero in the large-system limit within most (though not all) of the ferromagnetic phase. Here, we discuss how $\Delta p$ behaves for $P>3$.

We thus consider $P$ equally spaced marked players on a ring of $N > P$ qubits, and fix $P < \infty$ but set the spacing $l_P = N/P \to \infty$ in the large-system limit. In this limit, the cluster decomposition property of local magnetization operators in the Ising ground state\cite{pfeuty1970one} $\langle \hat{Z}_{i_j} \hat{Z}_{i_k}\rangle_0 \to \langle \hat{Z}_{i_j} \rangle_0 \langle \hat{Z}_{i_k} \rangle_0$, combined with translation invariance of this state yields
\begin{equation}
\label{eq:MFmanyplayers}
p_{\mathrm{qu}}(|\psi_0\rangle) = \frac{1}{2}\left(1 + \left(\frac{1+m}{2}\right)^P +  \left(\frac{1-m}{2}\right)^P \right),
\end{equation}
where $m = |\langle \psi_0 | \hat{Z}_i | \psi_0\rangle|$ denotes the local magnetization in the usual thermodynamic limit.

Thus, by playing a ``dilute'' version of the parity game, in which the marked players are separated by a distance $l_P \rightarrow \infty$ in the large-system limit, we have obtained a probability of success for the BBT protocol that depends only on the conventional order parameter for the ferromagnetic phase. In this sense, taking the limit of large $l_P$ for a nonlocal game is analogous to ``coarse-graining'' the underlying state\cite{cardy1996scaling}, because in this limit the players can only exploit long-range entanglement within their shared quantum state.

Equation \eqref{eq:MFmanyplayers} demonstrates that at a fixed finite $P$, the BBT protocol yields a non-infinitesimal quantum advantage over the optimal classical strategy
\begin{equation}
\Delta p = p_{\mathrm{qu}}(|\psi_0\rangle) - p_{\mathrm{cl}}^* > 0
\end{equation}
within most of the ferromagnetic phase, even in the large-system limit. However, Eq. \eqref{eq:MFmanyplayers} also implies that $\Delta p$ decreases rapidly as $m \to 0$ near the critical point.  In other words, the level of quantum advantage is sensitive not only to whether the system is in the ferromagnetic phase, but also to the magnitude of the ferromagnetic order parameter. For example, choosing $P = 8$ we find that $\Delta p \approx 0.27$ for $g=0.75$ and $\Delta p \approx 0.03$ for $g=0.99$, regardless of the value of $N \gg P$.

Let us now consider increasing $P$. We note that for fixed $1>m>\sqrt{2}-1$, $\Delta p > 0$ decays exponentially with $P$ as $P \to \infty$, while the threshold value of the transverse field strength $g_*$ at which quantum advantage is lost increases with $P$. In particular, in the large-system limit as $P \to \infty$, there is a battle of exponentials just as in Section \ref{sec:boe}, and quantum advantage is lost at precisely the mean-field threshold Eq. \eqref{eq:MFmagcondition} for the local magnetization.

\subsubsection{Demarcating the ferromagnetic phase}

The previous discussion raises the question of how near the critical point one can obtain a quantum advantage as $P  \to \infty$ in the $P$-bit parity game. To study this, we consider a double large-system limit with $N \gg P \gg 1$. As discussed above, in this limit quantum advantage is lost at the mean-field threshold Eq. \eqref{eq:MFmagcondition} for the magnetization. Using the known exact result\cite{pfeuty1970one} for the ground state magnetization $m = (1-g^2)^{1/8}$ of the TFIM as a function of the coupling strength, we deduce that quantum advantage is lost for
\begin{equation}
g \geq g_* = \sqrt{408\sqrt{2} - 576} \approx 0.9996.
\end{equation}
Thus we have succesfully demarcated $99.96\%$ of the ferromagnetic phase of the Ising model by using the $P$-bit parity game and the BBT protocol. This leaves the question of whether we can obtain a quantum advantage for some game in the remaining $0.04\%$ of the phase, and thereby use a nonlocal game and an appropriate protocol to characterize the ferromagnetic phase exactly.

We can achieve this in an asymptotic sense by reweighting the probability distribution of input bits for the $P$-bit parity game. Specifically, we now suppose that the $P$ input bits for the marked players are drawn from a Bernoulli distribution subject to the promise that $\sum_{j=1}^P a_{i_j}$ is even, i.e. bit strings $(a_{i_1},a_{i_2},\ldots,a_{i_P})$ consistent with the promise are drawn with probability 
\begin{equation}
p(\vec{a}) = 2\frac{\alpha^{\sum_{j=1}^P a_{i_j}} (1-\alpha)^{P - \sum_{j=1}^P a_{i_j}}}{1+(1-2\alpha)^P}
\end{equation}
where the parameter $\alpha \in (0,1)$, while all other input bits are set to $a_i = 0$.  For $\alpha < 0.5$ ($\alpha > 0.5)$, this biases the choices of $\vec{a}$ towards bit strings with $\frac{1}{P} \sum_{j=1}^P a_{i_j}$ close to $0$ $(1)$.

This reweighting modifies the expected probability of winning for the optimal classical strategy. We can calculate the latter by adapting previous results for the standard parity game\cite{brassard2005recasting}, and find that the advantage over random guessing for the optimal classical strategy is given by
\begin{equation}
\label{eq:alphaclass}
p_{\mathrm{cl}}^* - \frac{1}{2} = \begin{cases}
\frac{(\alpha^2 + (1-\alpha)^2)^{P/2}}{1+(1-2\alpha)^P}, & P \, \mathrm{even} \\
\frac{(\alpha^2 + (1-\alpha)^2)^{ (P-1)/2}(1-\alpha)}{1+(1-2\alpha)^P}, & P \, \mathrm{odd}
\end{cases},
\end{equation}
which recovers Eq. \eqref{eq:classwinprob} when $\alpha=1/2$. Subject to this distribution of inputs, the BBT protocol applied to the TFIM ground state, after the limits $N \to \infty$ and $l_P \to \infty$ are taken successively, yields a quantum probability of winning
\begin{equation}
p_{\mathrm{qu}}(|\psi_0\rangle) = \frac{1}{2}+ \frac{(1-\alpha + \alpha m)^P +(1-\alpha - \alpha m)^P}{2(1+(1-2\alpha)^P)}.
\end{equation}
This quantum strategy loses the battle of exponentials to the optimal classical strategy when
\begin{equation}
m \leq m_* = \frac{\sqrt{\alpha^2+(1-\alpha)^2}-(1-\alpha)}{\alpha}.
\end{equation}
As $\alpha \to 0^+$, we have
\begin{equation}
\label{eq:magthresh}
m_* = \frac{\alpha}{2} + \mathcal{O}(\alpha^2).
\end{equation}
It follows that quantum advantage is lost for coupling strengths
\begin{equation}
g \geq g_* = 1 - \frac{1}{512}\alpha^8 + \mathcal{O}(\alpha^{9}).
\end{equation}
This threshold can be made arbitrarily close to the quantum critical point as $\alpha \to 0^+$. However, while this game is a nonlocal game for $\alpha \in (0,1)$, the limiting game with $\alpha=0$ is not a true nonlocal game, as it is won with certainty by the deterministic classical strategy where all players always output $b_i=0$. In this sense, the property of being a nonlocal game is not continuous at $\alpha=0$. An analogous statement holds at the point $\alpha =1$. Thus we have constructed a family of nonlocal games parameterized by $\alpha \in (0,1)$, for which the BBT protocol applied to the Ising ground state asymptotically yields quantum advantage in precisely the ferromagnetic phase $|g| <1$ in the one-sided limit $\alpha \to 0^+$. 

Notice that analogous results hold for spin-$1/2$ quantum Ising models in any spatial dimension $d \geq 1$, since the threshold for losing quantum advantage in this construction, Eq. \eqref{eq:magthresh}, depends solely on the onsite magnetization in the large-system limit.

\section{The toric code topological phase}

In a companion paper\cite{companion}, we introduced the toric code game and its perfect quantum strategy, which wins the game with certainty when the players share the ground state of the ideal, fixed-point toric code Hamiltonian introduced by Kitaev\cite{KITAEV20032}.

We now turn to the question of whether the fixed-point protocol for the toric code game yields a quantum advantage in the toric code topological phase. We thus consider a more general Hamiltonian
on $2N$ bonds $b$ of a square lattice $\mathcal{L}$ with periodic boundary conditions in both directions:
\begin{equation}
\label{eq:TCH}
\hat{H} = - K \sum_{p} \hat{A}_p - K' \sum_s \hat{B}_s - h_X \sum_{b} \hat{X}_b - h_Z \sum_b \hat{Z}_b,
\end{equation}
where plaquette and star operators are defined as usual\cite{KITAEV20032,LaumannKitaev} by $\hat{A}_p = \prod_{b \in \partial p} \hat{Z}_b$ and  $\hat{B}_s = \prod_{b \in s} \hat{X}_b$. 
The ideal toric code Hamiltonian is obtained from Eq. \eqref{eq:TCH} by taking $h_X = h_Z = 0$. 

\subsection{The toric code game}

We begin by briefly recalling the the simplest version of the toric code game\cite{companion}. The toric code game has $2N$ players, with each player residing on a distinct bond of the lattice $\mathcal{L}$, and a referee or ``verifier'', who runs the game. At the beginning of the game, the verifier assigns $T \geq 3$ teams of players to vertical loops $\{\Gamma_i\}_{i=1}^T$ of the lattice. These players may communicate classically with other players within their vertical loop, but not with any other bonds of the lattice. Another set of players is assigned to a horizontal dual loop $\widetilde{\Gamma}$ of the lattice, which intersects each of the $\Gamma_i$ in exactly one bond. Players on the bonds of $\widetilde{\Gamma}$ may not communicate classically with one another. 

The toric code game then proceeds as follows. The verifier first provides each team $i$ with an input bit $a_i \in \{0,1\}$. Each player on a bond $b \in \widetilde{\Gamma}$ must then output a bit $y_b \in \{0,1\}$. The players collectively win the game if their outputs satisfy the condition
\begin{equation}
\sum_{b \in \widetilde{\Gamma}} y_b \equiv \frac{\sum_{i=1}^T a_i}{2} \, \mathrm{mod} \, 2.
\end{equation}
One can argue\cite{companion} that subject to the constraints above, the probability of winning for the optimal classical strategy is given by
\begin{equation}
\label{eq:tccwp}
p_{\mathrm{cl}}^* = \frac{1}{2} + \frac{1}{2^{\lceil T/2 \rceil}}.
\end{equation}

We now turn to the perfect quantum strategy for the toric code game. For the version of this game described above, this strategy uses the macroscopically entangled ground state
\begin{equation}
\label{eq:topcat}
|\psi_{\mathrm{TC}} \rangle = \frac{1}{\sqrt{2}} \left(|00\rangle + |0 1\rangle\right)
\end{equation}
of the ideal toric code, where
\begin{equation}
|0 0 \rangle = \mathcal{N} \prod_{s} (1+\hat{B}_s)\bigotimes_{b} |\hat{Z}_b = 1\rangle,
\end{equation}
$\mathcal{N}$ is a normalization constant, and the four degenerate ground states $|jk\rangle$, $j,k=0,1$ on the torus are labelled by their eigenvalues
\begin{equation}
\hat{W}_x |jk\rangle = (-1)^j |jk\rangle, \quad \hat{W}_y |jk\rangle = (-1)^k |jk\rangle
\end{equation}
with respect to Wilson loop operators $\hat{W}_{x/y} = \prod_{b \in \Gamma_{x/y}} \hat{Z}_b$, and can be obtained from the state $|00\rangle$ by acting with dual Wilson loop operators $\hat{V}_{x/y} = \prod_{b \in \widetilde{\Gamma}_{x/y}} \hat{X}_b$, so that $|jk\rangle = (\hat{V}_y)^j(\hat{V}_x)^k|00\rangle$.

The perfect quantum strategy then proceeds as follows. All players first share the entangled state $|\psi_{\mathrm{TC}}\rangle$. They then apply the following ``fixed-point protocol'', which we denote $\mathcal{P}_{\mathrm{TC}}$:
\begin{enumerate}
    \item Team $j$ acts with a nonlocal square root of their Wilson loop operator,
    \begin{equation}
    \hat{W}_j^{a_j/2} = \left(\frac{1+\hat{W}_j}{2}\right) + i^{a_j/2}\left(\frac{1-\hat{W}_j}{2}\right)
    \end{equation}
    where $\hat{W}_j = \prod_{b \in \Gamma_j} \hat{Z}_b$.
    \item Each player on a bond $b \in \widetilde{\Gamma}$ of the dual loop measures $\hat{X}_b$ to obtain the outcome $(-1)^{y_b}$ with $y_b \in \{0,1\}$, and returns the bit $y_b$.
\end{enumerate}
We refer to the companion paper\cite{companion} for an explanation of why this defines a perfect quantum strategy, together with a discussion of the many possible generalizations of the toric code game as described above.

\subsection{Effect of non-zero magnetic fields}
Let us now consider moving away from the ideal toric code ground state by introducing non-zero perturbing fields $h_X, \, h_Z \neq 0$. As for the Ising model discussed above, we seek to both (i) determine how far the fixed-point protocol for the toric code game yields a quantum advantage in the toric code phase and (ii) investigate whether this fixed-point protocol can be used to \emph{diagnose} the toric code phase. In contrast to our results for the Ising model, we will argue that the fixed-point protocol cannot yield a quantum advantage for the toric code game at generic values of $h_X$ and $h_Z$ within the toric code phase, and therefore cannot diagnose the toric code phase in any suitably defined thermodynamic limit, even when the teams are sufficiently well-separated in space that they are sensitive only to long-range entanglement within the ground state.

This failure is a consequence of the fact that the expectation values of the Wilson loop operators involved in playing the toric code game satisfy a ``perimeter law'' at generic points of the toric code phase. This means that the expectation value of a given Wilson loop decreases exponentially with the length of the loop in question. It follows that in the large-system limit, the expectation values of non-contractible Wilson loop operators, whose lengths are bounded below by the linear dimensions of the system, are generically vanishingly small, which ultimately destroys quantum advantage of the fixed-point protocol even far away from the critical region. One can evade this scenario to some extent if one of the perturbing fields $h_X, \, h_Z$ is set to zero; we will discuss this possibility in more detail below.

To justify the claims above, we first let $|\psi_0\rangle$ denote the  ground state of the Hamiltonian Eq. \eqref{eq:TCH} at some non-zero values of $h_X, h_Z$. As in the previous section, we consider the success rate of the quantum strategy $\mathcal{S} = (|\psi_0\rangle, \mathcal{P}_{\mathrm{TC}})$, which consists of applying the fixed-point protocol $\mathcal{P}_{\mathrm{TC}}$ to states away from the fixed point of the phase. For concreteness, suppose that there are $T$ teams of players on vertical loops $\Gamma_1, \, \Gamma_2,\ldots, \Gamma_T$. Using the identity \begin{equation}
\hat{W}_j^{-1/2} \hat{V}_{\widetilde{\Gamma}} \hat{W}_j^{1/2} = i \hat{W}_j \hat{V}_{\widetilde{\Gamma}}
\end{equation}
and averaging uniformly over input bit strings, one finds that the quantum probability of winning in this setting is given by
\begin{align}
\nonumber &p_{\mathrm{qu}}(|\psi_0\rangle) = \frac{1}{2} \\
\label{eq:TCexact}
+ &\frac{1}{2^T} \sum_{\{\vec{a} \in \{0,1\}^T : \sum_{i=1}^T a_i \, \mathrm{even}\}} \langle \psi_0 | \prod_{i=1}^T \hat{W}_i^{a_i} \hat{V}_{\widetilde{\Gamma}} | \psi_0\rangle.
\end{align}

To calculate the probability of winning away from the fixed point of the phase at second order in $h_X$ and $h_Z$, it suffices to consider the terms
\begin{align}
\nonumber |\psi_0\rangle \approx &\left(1-N\left(\frac{h_X}{4K}\right)^2 - N\left(\frac{h_Z}{4K'}\right)^2\right)\\
&\left(|\psi_{\mathrm{TC}}\rangle + \frac{h_X}{4K}\sum_b \hat{X}_b |\psi_{\mathrm{TC}}\rangle + \frac{h_Z}{4K'}\sum_b \hat{Z}_b |\psi_{\mathrm{TC}}\rangle\right)
\end{align}
in the perturbative expansion of $|\psi_0\rangle$ about $h_X=h_Z=0$ (one finds that the remaining second-order terms do not contribute to the calculation).

Computing expectation values in this perturbative approximation yields the simplification
\begin{align}
    \nonumber &p_{\mathrm{qu}}(|\psi_0\rangle) \approx \frac{1}{2} \\
+ &\frac{1}{2^T} \langle \hat{V}_{\widetilde{\Gamma}} \rangle_0 \sum_{\{\vec{a} \in \{0,1\}^T : \sum_{i=1}^T a_i \, \mathrm{even}\}} \prod_{i=1}^T \langle \hat{W}_i  \rangle_0^{a_i}
\end{align}
with\footnote{In order to obtain a non-zero result, the expectation value of a single Wilson loop $\langle \hat{W}_i \rangle_0$ must be taken in a non-macroscopically-entangled state, which is well-defined in the infinite-system limit, cf. the discussion in the introduction.}
\begin{align}
\label{eq:pertWilsonloop}
\langle \hat{V}_{\widetilde{\Gamma}} \rangle_0 \approx e^{-2L_x\left(\frac{h_Z}{4K'}\right)^2}, \,
\langle \hat{W}_{i} \rangle_0 \approx e^{-2L_y\left(\frac{h_X}{4K}\right)^2},
\end{align}
where $L_x$ and $L_y$ denote the horizontal and vertical extent of the lattice $\mathcal{L}$ respectively, so that the number of sites $N=L_xL_y$. Eq. (\ref{eq:pertWilsonloop}) reflects the fact that the two types of Wilson lines exhibit perimeter-law scaling in the toric code phase; although the calculation above is perturbative, such perimeter-law scaling is well-known to persist throughout the entirety of the topological phase\cite{gregor2011diagnosing}.

These expressions immediately suggest that if the number of teams $T$ is kept fixed, the strategy $\mathcal{S}$ loses quantum advantage in the large-system limit $L_x, L_y \to \infty$. Note that this holds true regardless of the spatial separation between the teams. One might ask whether there is some way to improve the quantum probability of winning for finite $T$, for example, by keeping one of the linear dimensions bounded or setting one of the perturbing magnetic fields to zero. We now explore these possibilities in more detail.

\subsubsection{The case $h_Z=0$,\, $L_y < \infty$}
First suppose that we set $h_Z = 0$ in Eq. \eqref{eq:TCH} and fix $L_y$ at some finite value. In this case, the expectation value $\langle \hat{V}_{\widetilde{\Gamma}}\rangle_0 = 1$ at all orders in perturbation theory, so we can make $L_x$ arbitrarily large without losing quantum advantage, and the quantum probability of winning is given by
\begin{equation}
p_{\mathrm{qu}}(|\psi_0\rangle) = \frac{1}{2} \\
+ \frac{1}{2^T} \sum_{\{\vec{a} \in \{0,1\}^T : \sum_{i=1}^T a_i \, \mathrm{even}\}} \langle \psi_0 | \prod_{i=1}^T \hat{W}_i^{a_i} | \psi_0\rangle.
\end{equation}

In particular, we can construct a game analogous to the $P$-bit parity game by fixing the number of teams $T$ and taking the spacing between the teams $l_T \to \infty$. Then, assuming that products of even numbers of vertical Wilson loops separated by a distance $l_T$ satisfy cluster decomposition,
\begin{equation}
\langle \psi_0 | \prod_{i=1}^T \hat{W}_i^{a_i} | \psi_0\rangle \to \prod_{i=1}^T \langle \psi_0 |  \hat{W}_i^{a_i} | \psi_0\rangle, \quad l_T \to \infty,
\end{equation}
(this is expected because the vertical Wilson loops are local operators for $L_y < \infty$), and using translation invariance, the quantum probability of winning can be written as
\begin{equation}
p_{\mathrm{qu}}(|\psi_0\rangle) = \frac{1}{2} \left(1 + \left(\frac{1+W }{2}\right)^T +  \left(\frac{1-W}{2}\right)^T \right),
\end{equation}
where $W =|\langle \psi_0 |  \hat{W}_i| \psi_0\rangle|$ in the $L_x = \infty$ ground state. Note that this formula is entirely analogous to Eq. \eqref{eq:MFmanyplayers}.

We see that the fixed-point protocol always yields a quantum advantage for the toric code game for sufficiently large $W$. As $T \to \infty$, the fixed-point protocol loses quantum advantage whenever the expectation values of individual Wilson loops
\begin{equation}
W \leq \sqrt{2} -1.
\end{equation}
Substituting in the perturbative parameter dependence of this expectation value according to Eq. \eqref{eq:pertWilsonloop}, we estimate that quantum advantage can survive within a range of perturbing fields
\begin{equation}
|h_X| \lesssim 4K \sqrt{\frac{|\log(\sqrt{2}-1)|}{2L_y}}
\end{equation}
that vanishes as $L_y \to \infty$.

\subsubsection{The case $h_X=0$, \, $L_x < \infty$}
Let us next consider the case $h_X = 0$. Now the expectation values of vertical Wilson loops $\langle \hat{W}_i\rangle_0 = 1$, so we may set $L_y$ to be arbitrarily large without losing quantum advantage. In this case, the quantum probability of winning simplifies to
\begin{equation}
p_{\mathrm{qu}}(|\psi_0\rangle) = \frac{1}{2}\left(1 + \langle \psi_0 | \hat{V}_{\widetilde{\Gamma}} | \psi_0\rangle\right).
\end{equation}
Thus the threshold for losing quantum advantage in this case is given simply by
\begin{equation}
\langle \psi_0 | \hat{V}_{\widetilde{\Gamma}} | \psi_0\rangle \leq \frac{1}{2^{\lceil T/2 \rceil-1}}.
\end{equation}
One can obtain an intuition for this result by considering a large number of teams $T \gg 1$ spaced equally along the horizontal extent $L_x$ of the torus. Then the perturbative result Eq. \eqref{eq:pertWilsonloop} implies that quantum advantage is lost whenever the distance between neighbouring teams becomes too large, i.e. when
\begin{equation}
l_T = \frac{L_x}{T} \gtrsim 4 \log{2} \left(\frac{K'}{h_Z}\right)^2.
\end{equation}
\subsubsection{The generic case revisited with $L_x, \, L_y < \infty$}
Finally, let us allow for non-zero perturbing fields $h_X, \, h_Z$ but keep the system dimensions finite, $L_x, \, L_y <\infty$. Combining the above results implies that quantum advantage can only survive for non-zero field strengths provided that
\begin{align}
|h_Z| \lesssim \frac{1}{L_x^{1/2}}, \quad |h_X| \lesssim \frac{1}{L_y^{1/2}},
\end{align}
which is a direct consequence of the perimeter-law scaling in Eq. \eqref{eq:pertWilsonloop}. We note that changing the configurations of teams and dual loops from our present convention\cite{companion} for the toric code game (vertical teams, horizontal dual loop) will modify these results accordingly.

To summarize, we have argued that for fixed $L_x$ and $L_y$, quantum advantage can survive over a small but non-zero range of perturbing fields $h_X$ and $h_Z$, while in the large-system limit $L_x, L_y \to \infty$, the fixed-point protocol fails to yield a quantum advantage for any $h_X, h_Z \neq 0$ regardless of the number of teams $T$, their orientation, and their relative spacing.

One way to understand the difference between the toric code and Ising models from the viewpoint of playing nonlocal games is that for the Ising model at any finite $N$, the ground states explicitly preserve the global $\mathbb{Z}_2$ symmetry; the macroscopic entanglement that provides a resource for winning the parity game is a result of the true ground state being an equal amplitude superposition of two states in distinct symmetry-broken sectors. The topological order of the toric code can similarly be viewed as arising from a spontaneously broken 1-form symmetry\cite{GKSW}.  However, for non-zero $h_X$ and $h_Z$, this 1-form symmetry is not exact, which leads to the perimeter-law behaviour of Wilson loops. 

\section{The $\mathbb{Z}_2 \times \mathbb{Z}_2$ SPT phase}
The multiplayer triangle game\cite{Cluster,Bravyi,DanielMiyake} is a nonlocal game that can be won with certainty using the ground state $|\psi_{\mathrm{CS}}\rangle$ of the $2N$-qubit stabilizer Hamiltonian
\begin{equation}
\hat{H} = -\sum_{i=1}^{2N} \hat{Z}_{i-1} \hat{X}_{i} \hat{Z}_{i+1}
\end{equation}
on a ring, with $\hat{Z}_{i-1}\hat{X}_{i}\hat{Z}_{i+1}|\psi_{\mathrm{CS}}\rangle = |\psi_{\mathrm{CS}}\rangle$. This cluster state can be viewed as a renormalization-group fixed point of the $\mathbb{Z}_2 \times \mathbb{Z}_2$ SPT phase in one dimension\cite{Verstraete,DanielMiyake}.

However, the triangle game and its multiplayer generalizations\cite{Bravyi,DanielMiyake} have an optimal classical probability of winning $p_{\mathrm{cl}}^* = 7/8$ that does not vary with the total number of players $N \geq 3$, essentially because the number of marked players $P$ who can receive non-zero input bits is fixed to $P=3$.

In Section \ref{sec:Pbitparity}, it was necessary to take the limit $P \to \infty$ to relate the parity game to a phase of matter. We thus introduce a $P$-bit generalization of the triangle game, which we call the ``polygon game''. This is a nonlocal game that admits a perfect quantum strategy using the cluster state $|\psi_{\mathrm{CS}}\rangle$ on $2P$ qubits. We then extend this game to a ``multiplayer polygon game'' with $N \geq 3$ players and $3 \leq P < N$ marked players, that admits a perfect quantum strategy $\mathcal{S}_{\mathrm{CS}} = (|\psi_{\mathrm{CS}}\rangle,\mathcal{P}_{\mathrm{CS}})$ using a $2N$-qubit cluster state. This game bears the precisely the same relation to the polygon game as the multiplayer triangle game does to the triangle game.

In contrast to the parity game\cite{brassard2005recasting}, we find that the polygon games become \emph{easier} to win classically as the number of players $P$ increases. We use this result to argue that as $P \to \infty$, the quantum strategy $\mathcal{S} = (|\psi\rangle, \mathcal{P}_{\mathrm{CS}})$ that uses a state $|\psi\rangle$ in the $\mathbb{Z}_2 \times \mathbb{Z}_2$ SPT phase only outperforms the classical result at the fixed point $|\psi\rangle = |\psi_{\mathrm{CS}}\rangle$.

\subsection{The polygon games}
\subsubsection{Rules of the polygon game}
\begin{figure}[t]
    \centering
    \includegraphics[width=0.95\linewidth]{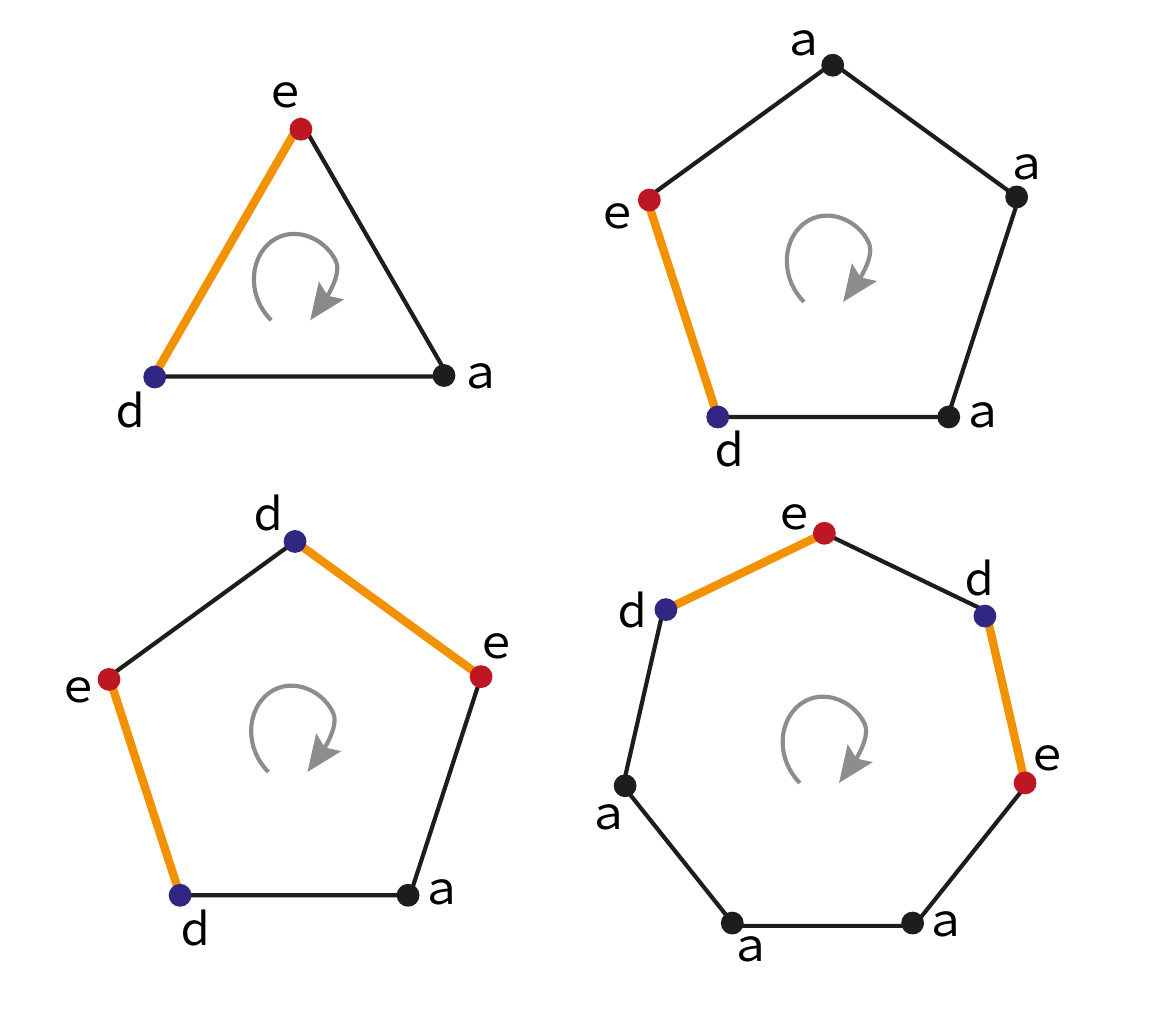}
    \caption{\textit{Top row:} Illustration of a single triangle condition in the polygon game with $P=3$ players (\textit{left}) and with $P=5$ players (\textit{right}). \textit{Bottom row:} Illustration of a single pentagon condition in the polygon game with $P=5$ players (\textit{left}) and $P=7$ players (\textit{right}). In each picture, the vertices that make up each edge $(i,i+1)$ of the appropriate matching are labelled with $d_i$ and $e_{i+1}$ respectively, according to the specified (clockwise) orientation. All other vertices $i$ are labelled with $a_i$.}
    \label{Fig3}
\end{figure}

The ``polygon game'' is a scalable family of $P$-player nonlocal games with $P \geq 3$, and is defined as follows. The $P$ players may not communicate classically, and each player receives an input bit $x_i \in \{0,1\}$. Player $i$ must output a bit string $(a_i,b_i,c_i)$ if they receive an input $x_i=0$ and a bit string $(d_i,b_i,e_i)$ if they receive an input $x_i=1$. In order for the players to win the game, they must satisfy the following conditions (which directly generalize those for the triangle game):
\begin{enumerate}
    \item If $x_i=0$, then $a_i+b_i+c_i$ must be even. If $x_i=1$, then $d_i + b_i + e_i$ must be even.
    \item For each input, $\sum_{i=1}^P b_i$ must be even. 
    \item If $\vec{x} = (0,0,\ldots,0)$, then $\sum_{i=1}^P a_i$ must be even.
    \item \emph{The ``triangle condition''}: If $\vec{x} = (1,1,0,\ldots,0)$, or any of its cyclic permutations with $x_i=x_{i+1}=1$, then $d_i+e_{i+1} + \sum_{j\neq i,i+1} a_j$ must be odd.
\end{enumerate}
For $P=3$ this recovers the triangle game\cite{Bravyi,DanielMiyake}. Additional possible conditions arise for $P>3$, each of which corresponds to a ``matching'' of the $P$-gon, in the sense of graph theory\cite{diestel2010graph}. For example, the triangle condition corresponds to the set of 1-edge matchings of the $P$-gon, and there are $P$ such conditions in total. The next simplest condition is the ``pentagon condition'' that arises for $P \geq 5$, namely
\begin{enumerate}
\setcounter{enumi}{4}
    \item \emph{The ``pentagon condition''}: If $\vec{x} = (1,1,1,1,0,\ldots,0)$, or any of its permutations with $x_i=x_{i+1}=1$ and $x_{j}=x_{j+1}=1$ for some $j \notin \{i-1,i,i+1\}$, then $d_i+e_{i+1} + d_j + e_{j+1} + \sum_{k\neq i,i+1,j,j+1} a_k$ must be even.
\end{enumerate}
There is one pentagon condition for every $2$-matching of the $P$-gon, for a total of $\frac{P (P-3)}{2}$ pentagon conditions. See Fig. \ref{Fig3} for an illustration of the triangle and pentagon conditions.

More generally, we can associate a polygon condition to any $r$-matching as follows:
\begin{enumerate}
\setcounter{enumi}{5}
    \item \emph{The ``$(2r+1)$-gon condition''}: For any $r$-matching $M_r$ of the $P$-gon with $2 < r \leq \lfloor P/2 \rfloor$, for the input $\vec{x}$ such that $x_i = x_{i+1} = 1$ whenever the edge $(i,i+1) \in M_r$ and $x_i=0$ otherwise, we must have
    \begin{equation}
    \sum_{(i,i+1)\in M_r} d_i+e_{i+1} + \sum_{(j-1,j),(j,j+1) \notin M_r} a_j \equiv r \mod{2}.
    \end{equation}
\end{enumerate}
This extends conditions 4 and 5 above. The number of such conditions for each $r$ is the number of $r$-matchings of the $P$-gon. These numbers define the triangle of coefficients of the Lucas polynomials\cite{koshy2019fibonacci}, and are equal to
$|M_r| = \frac{P}{P-r}\begin{pmatrix}P-r \\ r\end{pmatrix}$.

\subsubsection{Perfect quantum strategy for the polygon game}
To describe a perfect quantum strategy for the polygon game, it will be useful to write $\hat{S}_i = \hat{Z}_{i-1}\hat{X}_i \hat{Z}_{i+1}$ for the stabilizers of the cluster state, and define global symmetry operators
\begin{equation}
\label{eq:oddstab}
\hat{U}_a = \prod_{i=1}^{P} \hat{S}_{2i-1} = \prod_{i=1}^{P} \hat{X}_{2i-1}
\end{equation}
and
\begin{equation}
\label{eq:evenstab}
\hat{U}_b = \prod_{i=1}^{P} \hat{S}_{2i} = \prod_{i=1}^{P} \hat{X}_{2i},
\end{equation}
and the stabilizers
\begin{equation}
\label{eq:edgestab}
\hat{U}_M = \left(\prod_{(i,i+1)\in M} \hat{S}_{2i} \right) \hat{U}_a
\end{equation}
for each matching $M$ of the $P$-gon. By construction, each such stabilizer leaves the cluster state $|\psi_{\mathrm{CS}} \rangle$ invariant. The perfect quantum strategy is then identical to that for the triangle game\cite{Bravyi,DanielMiyake} (although the conditions for winning the polygon game are more stringent than for the triangle game), and proceeds as follows. First, all players share the $2P$-qubit cluster state $|\psi_{\mathrm{CS}}\rangle$ before playing the game, such that player $i$ has access to the $(2i-1)$-th and the $(2i)$-th qubit. They then apply the following protocol $\mathcal{P}_{\mathrm{CS}}$ to their qubits:
\begin{enumerate}
    \item If player $i$ receives the input $x_i=0$, they measure the tuple of commuting operators $(\hat{X}_{2i-1},\hat{X}_{2i},\hat{X}_{2i-1}\hat{X}_{2i})$ on their two qubits, and return the bit string $(a_i,b_i,c_i)$ defined by the measurement outcome $((-1)^{a_i},(-1)^{b_i},(-1)^{c_i})$.
    \item If player $i$ receives the input $x_i=1$, they measure the tuple of commuting operators $(\hat{Y}_{2i-1}\hat{X}_{2i},\hat{X}_{2i},\hat{Y}_{2i-1})$ on their two qubits, and return the bit string $(d_i,b_i,e_i)$ defined by the measurement outcome $((-1)^{d_i},(-1)^{b_i},(-1)^{e_i})$.
\end{enumerate}
The outcomes of the full set of commuting measurements on $|\psi_{\mathrm{CS}}\rangle$ are constrained by the unit eigenvalues of the stabilizer operators Eqs. \eqref{eq:oddstab}-\eqref{eq:edgestab} on this state to always satisfy the conditions of the polygon game. Thus the polygon game admits the perfect quantum strategy $\mathcal{S}_{\mathrm{CS}} = (|\psi_{\mathrm{CS}}\rangle,\mathcal{P}_{\mathrm{CS}})$.

\subsubsection{Classical winning probabilities}
For an even number of players $P$, the polygon game can be won classically with probability one, and therefore does not define a nonlocal game. A perfect classical strategy is given by
\begin{equation}
(a_i,b_i,c_i) = (d_i,b_i,e_i) = (1,1,0)
\end{equation}
for all $i$. Thus the optimal classical winning probability
\begin{equation}
p^*_{\mathrm{cl}}(P) = 1, \quad P \, \mathrm{even}.
\end{equation}

For an odd number of players $P$, there is no longer a perfect classical strategy, because the triangle condition, Condition 4, contradicts Conditions 1, 2, and 3, as for the three-player game. Thus
\begin{equation}
p^*_{\mathrm{cl}}(P) \leq 1- \frac{1}{2^P}, \quad P \, \mathrm{odd},
\end{equation}
which implies that for odd $P \geq 3$, the polygon game is a true nonlocal game. For the triangle game, the upper bound is saturated and we have\cite{Bravyi} $
p^*_{\mathrm{cl}}(3) = 7/8$. For the pentagon game, we have proved by numerical exhaustion of all $2^{20} = 1048576$ possible classical deterministic strategies that
\begin{equation}
p^*_{\mathrm{cl}}(5) = 29/32.
\end{equation}
For larger odd $P$ we can achieve a useful lower bound on the classical probability of winning by considering the classical strategy
\begin{align}
\nonumber (a_i,b_i,c_i) &= (d_i,b_i,e_i) = (1,1,0), \quad 1 \leq i < P, \\
(a_P,b_P,c_P) &= (0,0,0), \, (d_P,b_P,e_P) = (1,0,1).
\end{align}
This strategy loses for all inputs with $x_1=x_P=1$. This includes one triangle condition, $P-3$ pentagon conditions, and generally $\begin{pmatrix} P-1-r \\ r-1\end{pmatrix}$ $(2r+1)$-gon conditions, for a total of
\begin{equation}
N_{\mathrm{losses}}(P) = \sum_{r=0}^{(P-3)/2} \begin{pmatrix} P -2 -r \\ r \end{pmatrix} = F_{P-1}
\end{equation}
losses, where $F_{P}$ denotes the $P$th Fibonacci number, indexed so that $F_1=F_2=1$. This implies a lower bound
\begin{equation}
\label{eq:classwinbound}
p_{\mathrm{cl}}^*(P) \geq \frac{2^P - F_{P-1}}{2^P} = 1 - \frac{F_{P-1}}{2^P}, \quad P \,\mathrm{odd}.
\end{equation}
We conjecture that this bound is tight. Note that it implies
\begin{equation}
p_{\mathrm{cl}}^*(P) \to 1, \quad P \to \infty,
\end{equation}
since
\begin{equation}
p_{\mathrm{cl}}^*(P) \geq 1 - \frac{F_{P-1}}{2^P} \sim 1 - \frac{1}{\phi \sqrt{5}}(\phi/2)^P \to 1, \quad P \to \infty,
\end{equation}
where $\phi = (1+\sqrt{5})/2$ denotes the Golden Ratio. Thus unlike the parity game and the toric code game, the polygon games cease to define nonlocal games in the limit of infinitely many players $P = \infty$.

\subsubsection{The multiplayer polygon game}
Finally we introduce a polygonal generalization of the multiplayer triangle game\cite{Bravyi}. Thus consider a $P$-gon inscribed within an $N$-gon with $N \geq P$ and $P$ odd, at vertices $i_1<i_2<\ldots < i_P$ of the $N$-gon. For each edge $(i_j,i_{j+1})$ of the inscribed $P$-gon, we let $E_{j} = \{i_j < i < i_{j+1}\}$ denote the corresponding sites of the outer $N$-gon. Then the rules of the ``$N$-player $P$-gon game'' are as follows (see Fig. \ref{Fig4} for a visualization):

\begin{figure}[t]
    \centering
    \includegraphics[width=0.99\linewidth]{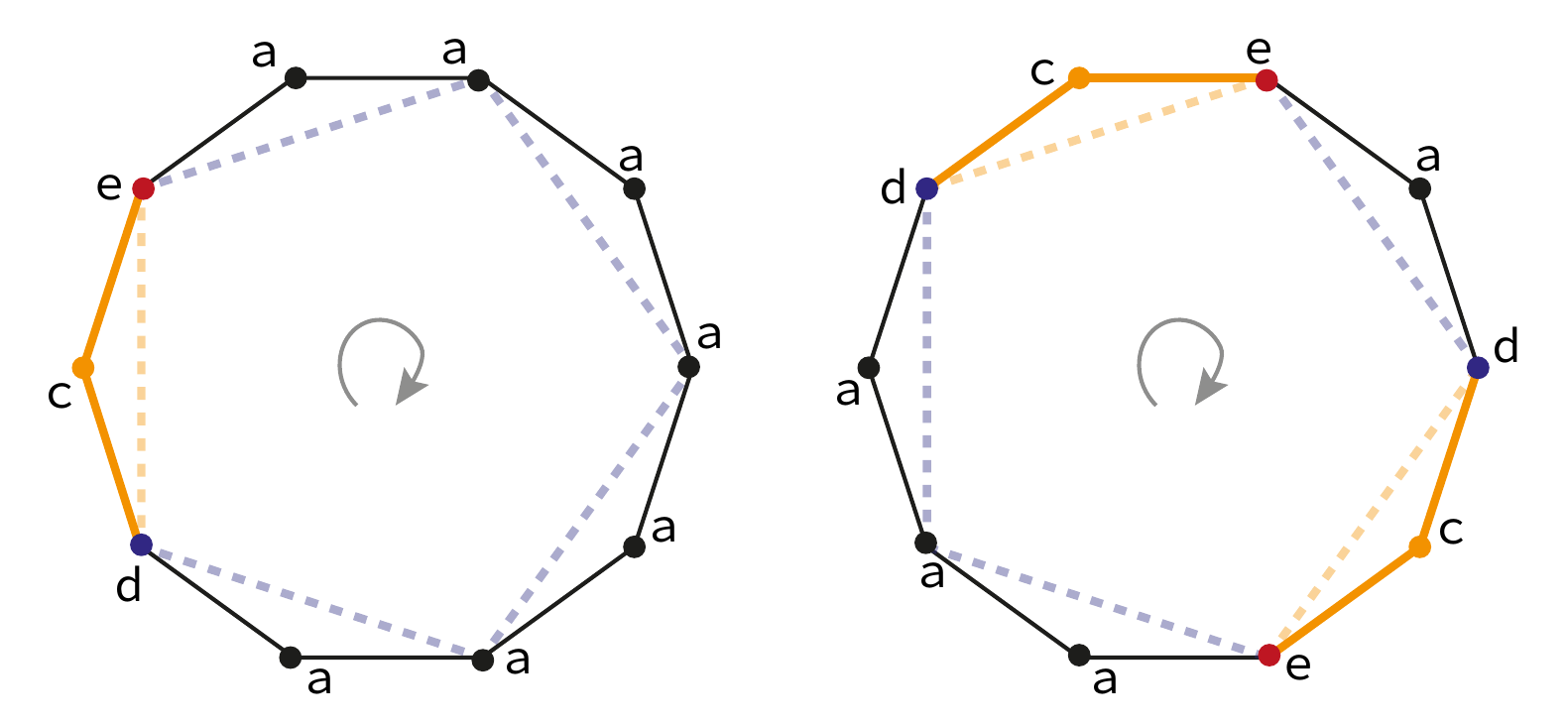}
    \caption{Illustration of the $N$-player $P$-gon game for $N=10$ and $P=5$, i.e. a pentagon inscribed within a decagon. We depict a single triangle condition (\textit{left}) and a single pentagon condition (\textit{right}). In each such picture, the two vertices that make up an edge $(i_j,i_{j+1})$ of the matching of the inscribed polygon are labelled by $d_{i_j}$ and $e_{i_{j+1}}$ respectively, according to the specified (clockwise) orientation. Vertices on the outer polygon that make up the set $E_j = {i_j < i < i_{j+1}}$ are labelled by $c_i$. All vertices of the outer polygon that are not involved in the matching of the inner polygon are labelled by $a_i$.}
    \label{Fig4}
\end{figure}

\begin{enumerate}
\setcounter{enumi}{-1}
    \item Players at the vertices of the inscribed $P$-gon can receive input bits $x_i \in \{0,1\}$. All other players receive an input bit $x_i=0$.
    \item If $x_i=0$, then $a_i+b_i+c_i$ must be even. If $x_i=1$, then $d_i + b_i + e_i$ must be even.
    \item For all inputs, $\sum_{i=1}^N b_i$ must be even. 
    \item If $\vec{x} = (0,0,\ldots,0)$, then $\sum_{i=1}^N a_i$ must be even.
    \item \emph{The ``$(2r+1)$-gon condition''}: For any $r$-matching $M_r$ of the inscribed $P$-gon with $1 \leq r \leq \lfloor P/2 \rfloor$, for the input $\vec{x}$ such that $x_{i_j} = x_{i_{j+1}} = 1$ whenever the edge $(i_j,i_{j+1}) \in M_r$ and $x_i=0$ otherwise, we must have
    \begin{align}
    \nonumber
    &\sum_{(i_j,i_{j+1})\in M_r} d_{i_j}+e_{i_{j+1}} \\+
    \nonumber &\sum_{(i_{k-1},i_k),(i_k,i_{k+1}) \notin M_r} a_{i_k}
    \equiv r + \sum_{(i_j,i_{j+1})\in M_r} \sum_{l\in E_j} c_l \\
    &+ \sum_{(i_{k-1},i_k),(i_k,i_{k+1}) \notin M_r} \sum_{l\in E_k} a_l \mod{2}.
    \end{align}
\end{enumerate}

This defines a nonlocal game for essentially the reasons discussed above for the polygon game, and its perfect quantum strategy is again given by $\mathcal{S}_{\mathrm{CS}} = (|\psi_{\mathrm{CS}}\rangle,\mathcal{P}_{\mathrm{CS}})$, where $|\psi_{\mathrm{CS}}\rangle$ is now the cluster state on $2N$ qubits.

Notice that the optimal classical probability of winning for the $N$-player $P$-gon game is equal to $p_{\mathrm{cl}}^*(P)$ for the $P$-player $P$-gon game, since the players off the inscribed $P$-gon can all return $(a_i,b_i,c_i) = (0,0,0)$ without diminishing the success of the optimal classical strategy on the $P$-gon. As for the $N$-player, $P$-bit parity game, this is true mathematically speaking whether or not the $N$ players are told which of them will be marked before the game is played (though there is a separate question of how well the players can coordinate in advance to reliably attain such an optimal strategy). To avoid this difficulty, we shall henceforth assume that the players know whether they are marked in advance.

\subsection{Playing polygon games in the SPT phase}
Let us now consider playing the $N$-player $P$-gon game with a $2N$ qubit state $|\psi_0\rangle$ that is in the $\mathbb{Z}_2 \times \mathbb{Z}_2$ SPT phase. This problem was previously studied in the case $P=3$, for which it was found that the quantum strategy $\mathcal{S} = (|\psi_0\rangle,\mathcal{P}_{\mathrm{CS}})$ could yield a quantum advantage for the multiplayer triangle game away from the cluster state fixed point as $N\rightarrow \infty$\cite{DanielMiyake}. Intuitively, this is possible because the three marked players are spatially well-separated; as we found for the Ising ground state, such spatial separation allows the marked players to achieve quantum advantage over a substantial portion of the phase. As for the $3$-bit parity game, however, the fixed-point protocol for the multiplayer triangle game loses quantum advantage at a non-zero value of the string order parameter diagnosing the SPT phase, and is therefore insufficient for clearly identifying the phase.

To determine whether this situation improves for $P>3$ marked players, we first note that the total number of $(2r+1)$-gon conditions for $1 \leq r \leq \lfloor P/2 \rfloor$ is given by the $P$th Lucas number, $L_P$, minus one, where $L_1 = 1$, $L_2=3$ and $L_P = L_{P-1}+L_{P-2}$. Thus the quantum probability of winning can be written (using the global $\mathbb{Z}_2$ symmetries in the SPT phase and the properties of dichotomic observables\cite{DanielMiyake}) as
\begin{align}
\nonumber p_{\mathrm{qu}}(|\psi_0\rangle) = &1-\frac{(L_P-1)}{2^{P+1}} \\
+ &\frac{1}{2^{P+1}}\sum_{r=1}^{\lfloor P/2 \rfloor} \sum_{r\mathrm{-matchings \,} M_r} \langle \psi_0| \hat{U}_{M_r} | \psi_0\rangle
\end{align}
where the stabilizers arising in the multiplayer polygon game
\begin{align}
\hat{U}_{M_r} = \prod_{\{j:(i_j,i_{j+1})\in M_r\}} \hat{S}_j
\end{align}
can be interpreted as products of string order parameters\cite{DanielMiyake}
\begin{equation}
\hat{S}_j = \hat{Z}_{2i_j-1}\hat{X}_{2i_j} \left(\prod_{l \in E_j} \hat{X}_{2l}\right)\hat{Z}_{2i_{j+1}+1},
\end{equation}
with one string order parameter for each edge of the inscribed $P$-gon.

To proceed further, let us fix $P < \infty$, assume that the vertices of the inscribed $P$-gon are evenly spaced, pass to the large-system limit $N \gg P$ and make the estimate\footnote{A strict equality here would be equivalent to the assumption that the expectation values of products of string operators whose regions of support are separated by an extensive distance satisfy the cluster decomposition property in the SPT state as $N \to \infty$. Note that this holds at the cluster-state fixed point $|\psi_0\rangle = |\psi_{\mathrm{CS}}\rangle$.}
\begin{equation}
\label{eq:stringcluster}
\langle \psi_0 | \hat{U}_{M_r} | \psi_0 \rangle \approx \langle \psi_0 | \hat{S} |  \psi_0\rangle^r,
\end{equation}
where $\hat{S}$ is the string order parameter corresponding to a single edge and is a product of $\sim N/P$ single-qubit operators for any finite $N$. Then, writing $\langle \hat{S} \rangle_0 = \langle \psi_0 | \hat{S} | \psi_0 \rangle$, we can estimate the quantum probability of winning as
\begin{equation}
p_{\mathrm{qu}}(|\psi_0\rangle) \approx 1-\frac{(L_P-1)}{2^{P+1}} + \frac{1}{2^{P+1}}\sum_{r=1}^{\lfloor P/2 \rfloor} |M_r| \langle \hat{S} \rangle_0^{r}.
\end{equation}
Note that the function of the order parameter appearing in this equation is (one definition of) the matching polynomial of the inscribed $P$-gon, and can be expressed in terms of the Lucas polynomials\cite{koshy2019fibonacci}
\begin{equation}
\label{eq:Lucas}
L_P(x) = \frac{1}{2^P}\left[(x+\sqrt{x^2+4})^P +(x-\sqrt{x^2+4})^P\right]
\end{equation}
as
\begin{equation}
\label{eq:qwp}
p_{\mathrm{qu}}(|\psi_0\rangle) \approx 1-\frac{L_P-\langle \hat{S} \rangle_0^{P/2}L_P (\langle \hat{S}\rangle_0^{-1/2} )}{2^{P+1}}.
\end{equation}

We expect that this strategy loses quantum advantage when our estimate for the quantum probability of winning Eq. \eqref{eq:qwp} falls below the classical lower bound Eq. \eqref{eq:classwinbound}. This happens at a threshold value of the string order parameter
\begin{equation}
\langle \hat{S} \rangle_0^{P/2}L_P (\langle \hat{S}\rangle_0^{-1/2}) \lesssim \left( L_P -  2 F_{P-1}\right).
\end{equation}
The corresponding threshold value of $\langle \hat{S} \rangle_0$ is an increasing function of $P$, which means that in contrast to our findings for the Ising model, there is no improvement gained in identifying the ordered phase by increasing $P$. In the large $P$ limit, this threshold becomes
\begin{equation}
\langle \hat{S} \rangle_0^{P/2}L_P (\langle \hat{S}\rangle_0^{-1/2}) \lesssim \left(1-\frac{2}{\phi \sqrt{5}}\right)\phi^P,
\end{equation}
where we have used the fact that $L_P \sim \phi^P$ as $P \to \infty$. The leading asymptotic behaviour of the left-hand side follows from the definition of the Lucas polynomials Eq. \eqref{eq:Lucas} and is given by
\begin{equation}
\langle \hat{S} \rangle_0^{P/2}L_P (\langle \hat{S}\rangle_0^{-1/2}) \sim \left(\frac{1 + \sqrt{1+4\langle \hat{S} \rangle_0}}{2}\right)^P.
\end{equation}
Thus we expect that this quantum strategy loses the battle of exponentials for sufficiently large $P$ whenever
\begin{equation}
\frac{1+\sqrt{1+4\langle \hat{S}\rangle_0}}{2} < \phi,
\end{equation}
which is true whenever
\begin{equation}
\langle \hat{S} \rangle_0 < 1.
\end{equation}
The above argument suggests that as $P \to \infty$, a state in the $\mathbb{Z}_2 \times \mathbb{Z}_2$ SPT phase only provides quantum advantage for the multiplayer $P$-gon game at the fixed point with $\langle \hat{S} \rangle_0 = 1$.

Finally, we note that unlike for the parity game (see Theorem \ref{thm:thm}) and the toric code game\cite{companion}, the condition that the quantum strategy $\mathcal{S} = (|\psi\rangle,\mathcal{P}_{\mathrm{CS}})$ is a perfect quantum strategy for the polygon game does not uniquely determine the state $|\psi\rangle$ to be the fixed-point cluster state $|\psi_{\mathrm{CS}}\rangle$, as it fixes only half the stabilizers of the state $|\psi_{\mathrm{CS}}\rangle$ in the large-system limit. This is a further indication that the polygon games (including the triangle game) are insufficient to capture the $\mathbb{Z}_2 \times \mathbb{Z}_2$ SPT phase.

\section{Conclusion}

We have systematically investigated the question of how far nonlocal games can be won using phases of quantum matter. Our results, in combination with other recent studies of playing nonlocal games with condensed matter ground states\cite{DanielMiyake,companion}, bring the foundational (but so far largely theoretical) notion of quantum pseudo-telepathy\cite{Brassard2005} closer to the kinds of entangled many-body wavefunctions that are realized in low-temperature condensed matter systems in the laboratory. From a practical perspective, our results for the quantum Ising model are closest to present-day experimental capabilities\cite{Coldea_2010}. However, if the protocols discussed in this paper for the Ising model are to be realized experimentally, it is important that the system lies close to its true ground state, i.e. is maintained at a temperature that is small compared to the gap separating the two lowest-lying eigenstates. If this is not the case, then mixing between these two opposite-parity states will eliminate any quantum advantage for the parity game gained by the BBT protocol.

Even at zero temperature, our findings reveal a new and possibly unexpected distinction between conventional symmetry-breaking phases and topological and SPT phases from the viewpoint of winning nonlocal games, with conventional symmetry-breaking phases apparently yielding a better resource for winning such games. One way to understand this distinction is that the games considered above admit quantum strategies whose quantum advantage depends continuously on the order parameters of an underlying phase of matter.

For example, we have proved that the ground state of the transverse-field Ising model in one dimension provides a quantum advantage for winning the parity game over the entirety of its ferromagnetic phase. We expect that similar conclusions hold for the ground states of $\mathbb{Z}_M$ clock models with $M>2$, which can be used to win the modulo $M$ generalizations of the parity game due to Boyer\cite{boyer2004extended}. (We prove some new results on these games in Appendix \ref{app:Boyer}, including an improved upper bound on the classical probability of winning and an analogue of Theorem \ref{thm:thm}.)

In contrast, it appears that topological and SPT phases do not tend to confer a quantum advantage for nonlocal games, beyond a region of the phase diagram that becomes vanishingly small in the large-system limit. For the deconfined ground states of the perturbed toric code Hamiltonian, this is because the expectation values of Wilson loop operators decay exponentially with the linear system size away from the ideal toric code limit, reflecting their perimeter-law scaling. For the $\mathbb{Z}_2 \times \mathbb{Z}_2$ SPT phase, the difficulty instead seems to be that the triangle game studied in previous work\cite{DanielMiyake} and its generalization to the ``polygon games'' introduced above are simply too easy to win classically, and therefore do not probe sufficient entanglement to determine the SPT phase. Both these observations are perhaps surprising compared to our results for the Ising model, given the widespread expectation that topological and SPT phases of matter encode quantum information more robustly than more conventional phases of quantum matter\cite{TopQCreview,SPTcomputer}. This might be a consequence of our choice of protocols $\mathcal{P}$, as we now discuss.

In this paper, we considered quantum strategies $\mathcal{S} = (|\psi\rangle,\mathcal{P})$ such that the state $|\psi\rangle$ could vary within a given phase of matter, but constrained the protocol $\mathcal{P}$ to be the ``fixed-point protocol'' for that phase. In several cases, this choice led to a substantial diminution of the quantum probability of winning away from the fixed point of the phase, in the limit of a large number of players $N \to \infty$. One might ask whether it is possible to perform substantially better by modifying the protocol $\mathcal{P}$, i.e. to find a quantum strategy that (i) uses ground states $|\psi_0\rangle$ in a given phase of matter, away from the $\xi=0$ fixed point and (ii) wins a nonlocal game $\mathcal{G}$ with probability near one, $p_{\mathrm{qu}} \approx 1$ as $N \to \infty$. Some indication that such protocols do exist comes from renormalization group ideas\cite{cardy1996scaling,Verstraete}; away from critical points and in the large-system limit, ground states have a finite correlation length $\xi < \infty$, and on length scales larger than this correlation length, quantum fluctuations should be suppressed so that the coarse-grained ground state resembles a state with $\xi=0$. This suggests that a non-critical ground state, whose total number of qubits $N$ is much larger than the correlation length $\xi$, could be used to win an $N' = N/\xi$-player version of $\mathcal{G}$ with probability $p_{\mathrm{qu}} \approx 1$, using a modification of the fixed-point protocol whereby each player acts on order $\xi$ qubits at a time. 

Indeed, for the ferromagnetic phase of the quantum Ising model, this idea is approximately realized by the ``$P$-bit parity games'' that we construct in Section \ref{sec:Pbitparity}, and for $P=3$ bits, we do find a quantum probability of winning that is approximately equal to one within a finite region of the ferromagnetic phase. However, if such a protocol could be constructed directly from local operators in the manner sketched above, it would require a more general formulation of nonlocal games than has been considered in the past\cite{Brassard2005}, because it must allow for classical communication between multiple qubits within a single correlation length. A fundamental problem with allowing for classical communication between neighbouring qubits is that it allows for classical communication between \emph{any} two qubits in the system, through successively passing messages between neighbours, and therefore eliminates any quantum advantage for these games. Previous work has circumvented this problem by only allowing a limited ``communication distance'' between qubits\cite{Cluster}, which is reasonable as a matter of principle but not especially physical. We expect that introducing a non-zero communication distance for the games discussed in this paper would both improve the effectiveness of the best classical strategies and allow for better quantum strategies away from the fixed points of phases, but leave a detailed consideration of such improvements to future studies.

This brings us to the closely related question of whether quantum games can be used to \emph{uniquely} characterize the ground states of condensed matter systems. To be precise, given a game $\mathcal{G}$ and a quantum protocol $\mathcal{P}$, one can ask whether the condition that $\mathcal{S} = (|\psi\rangle,\mathcal{P})$ is a perfect quantum strategy for $\mathcal{G}$ imposes non-trivial constraints on the state $|\psi\rangle$. Our Theorem \ref{thm:thm} proves that when the game $\mathcal{G}$ is the parity game and the protocol $\mathcal{P}$ is the Brassard-Broadbent-Tapp protocol, $\mathcal{S}$ is a perfect quantum strategy for $\mathcal{G}$ if and only if the state $|\psi\rangle$ equals $|\mathrm{GHZ}^+\rangle$ up to a global phase. Thus the $\xi=0$ fixed point of the ferromagnetic phase of the quantum Ising model is uniquely determined by the BBT protocol for the parity game. We have proved analogous results relating Boyer's modulo $M$ game to the quantum clock model (see Theorem \ref{thm:thm2}), and relating the toric code game to the toric code ground state\cite{companion}. 

The latter results might seem to rely on the fact that the underlying fixed-point protocols are equivalent to contextual families of measurements of stabilizers, which naturally give rise to nonlocal games\cite{Abramsky2017}. However, our results for the quantum Ising model in Section \ref{sec:Ising} demonstrate clearly that vestiges of contextuality can persist through entire phases of quantum matter, and in conjunction with specific protocols and specific nonlocal games, demarcate these phases sharply. Extending these observations to a full ``device-independent self-testing'' of phases of matter, without reference to any specific protocol\cite{SelfTest,DeviceIndependent,SelfTestVaz,MagicSelfTesting,Grilo}, poses a fascinating challenge for future work.

\section{Acknowledgments}
The authors thank I. Arad, D. S. Borgnia, A. B. Grilo, R. M. Nandkishore, and especially U. V. Vazirani for helpful discussions. V. B. B. is supported by a fellowship at the Princeton Center for Theoretical Science. F. J. B. is supported by NSF DMR-1928166, and is grateful to the Carnegie Corporation of New York and the Institute for Advanced Study, where part of this work was carried out. This work was supported by a Leverhulme Trust International Professorship grant number LIP-202-014 (S. L. S). For the purpose of Open Access, the author has applied a CC BY public copyright licence to any Author Accepted Manuscript version arising from this submission.
\bibliography{games.bib}

\clearpage
\onecolumngrid
\appendix

\section{Proof of Theorem \ref{thm:thm}}
\label{app:Theorem1}
In this Appendix, we prove Theorem \ref{thm:thm} on the probability of winning the parity game with the quantum strategy $\mathcal{S} = (|\psi\rangle,\mathcal{P}_{\mathrm{BBT}})$, with allowed input bit strings $\vec{a}$ selected uniformly at random. First suppose that the components of the (normalized) state $|\psi\rangle$ in the computational basis are given by
\begin{equation}
    |\psi\rangle = \sum_{\vec{\sigma} \in \{0,1\}^N} c_{\vec{\sigma}} | \sigma_1 \sigma_2\ldots \sigma_N \rangle.
\end{equation}
After Step 1, this becomes
\begin{equation}
\label{eq:trick}
|\psi'\rangle = \sum_{\vec{\sigma} \in \{0,1\}^N} i^{\sum_{j=1}^N \sigma_j a_j} c_{\vec{\sigma}}|\vec{\sigma}\rangle.
\end{equation}
After Step 2, we have
\begin{align}
\nonumber |\psi''\rangle = \frac{1}{2^{N/2}} \sum_{\vec{\sigma} \in \{0,1\}^N} i^{\sum_{j=1}^N \sigma_j a_j} c_{\vec{\sigma}} \bigotimes_{\{j:\sigma_j=0\}} (|0\rangle+|1\rangle)  \bigotimes_{\{j:\sigma_j=1\}} (|0\rangle-|1\rangle).
\end{align}
We can use the same notational trick as in Eq. \eqref{eq:trick} to write the tensor product as
\begin{align}
\bigotimes_{\{j:\sigma_j=0\}} (|0\rangle+|1\rangle)  \bigotimes_{\{j:\sigma_j=1\}} (|0\rangle-|1\rangle) = \sum_{\vec{b}\in \{0,1\}^N} (-1)^{\sum_{j=1}^N\sigma_j b_j} |b_1 b_2\ldots b_2 \rangle,
\end{align}
which yields the double sum
\begin{equation}
|\psi''\rangle = \frac{1}{2^{N/2}} \sum_{\vec{\sigma} \in \{0,1\}^N}\sum_{\vec{b}\in \{0,1\}^N} e^{i\pi \sum_{j=1}^N \sigma_j(a_j/2+b_j)} c_{\vec{\sigma}} |b_1 b_2\ldots b_2 \rangle.
\end{equation}
Interchanging the order of summation we can write this as
\begin{equation}
|\psi''\rangle = \sum_{\vec{b}\in \{0,1\}^N} \left(\frac{1}{2^{N/2}} \sum_{\vec{\sigma} \in \{0,1\}^N} e^{i\pi \sum_{j=1}^N \sigma_j(a_j/2+b_j)} c_{\vec{\sigma}}\right) |b_1 b_2\ldots b_2 \rangle,
\end{equation}
from which the probability of winning the parity game with input bit string $\vec{a}$ after Step 3 can be read off to be
\begin{equation}
p(|\psi\rangle,\vec{a}) = \frac{1}{2^{N}} \sum_{\{\vec{b}:\sum_{j=1}^N b_j \equiv r \,\mathrm{mod} \,2\}} \left|  \sum_{\vec{\sigma} \in \{0,1\}^N} e^{i\pi \sum_{j=1}^N \sigma_j(a_j/2+b_j)} c_{\vec{\sigma}}\right|^2
\end{equation}
by the Born rule, where $r = \sum_{j=1}^N a_j/2$ denotes the sought-after parity. Perhaps surprisingly, this sum admits drastic simplification. To see this, we first write
\begin{equation}
p(|\psi\rangle,\vec{a}) = \frac{1}{2^{N}}  \sum_{\vec{\sigma},\vec{\sigma}' \in \{0,1\}^N} c_{\vec{\sigma}}c_{\vec{\sigma}'}^*e^{i\pi\sum_{j=1}^N (\sigma_j-\sigma'_j)a_j/2} \sum_{\{\vec{b}:\sum_{j=1}^N b_j \equiv r \,\mathrm{mod} \,2\}} e^{i\pi \sum_{j=1}^N (\sigma_j-\sigma'_j)b_j}. 
\end{equation}
The second sum can be performed using the identity
\begin{equation}
\label{identity}
\sum_{\{\vec{b}:\sum_{j=1}^N b_j \equiv r \,\mathrm{mod} \,2\}} \prod_{j=1}^N z_j^{b_j} = \frac{1}{2} \left( \prod_{j=1}^N (1+z_j) + (-1)^r  \prod_{j=1}^N (1-z_j)\right),
\end{equation}
which yields
\begin{align}
\sum_{\{\vec{b}:\sum_{j=1}^N b_j \equiv r \,\mathrm{mod} \,2\}} e^{i\pi \sum_{j=1}^N (\sigma_j-\sigma'_j)b_j} = 2^{N-1}\left[\prod_{j=1}^N \delta_{\sigma_j,\sigma'_j} + (-1)^r \prod_{j=1}^N \delta_{\sigma_j,1-\sigma'_j} \right],
\end{align}
implying that
\begin{equation}
p(|\psi\rangle,\vec{a}) = \frac{1}{2}\left(1+ \sum_{\vec{\sigma}\in \{0,1\}^N} e^{i\pi \sum_{j=1}^N \sigma_j a_j}c_{\vec{\sigma}}c^*_{\vec{1}-\vec{\sigma}}\right),
\end{equation}
where it is useful to define fully polarized bit-strings $\vec{0} = (0,0,\ldots,0)$ and $\vec{1} = (1,1,\ldots,1)$. To obtain the probability of winning the parity game for a uniformly random input bit string, we must average over all allowed input bit strings, $\vec{a} \in \{0,1\}^N$ with $\sum_{j=1}^N a_j$ even. This yields the quantum winning probability
\begin{equation}
p_{\mathrm{qu}}(|\psi\rangle)  = \frac{1}{2^{N-1}}\sum_{\{\vec{a}:\sum_{j=1}^N a_j \, \mathrm{even}\}} p(|\psi\rangle,\vec{a}) = \frac{1}{2} + \frac{1}{2^N} \sum_{\{\vec{a}:\sum_{j=1}^N a_j \, \mathrm{even}\}} \sum_{\vec{\sigma}\in \{0,1\}^N} e^{i\pi \sum_{j=1}^N \sigma_j a_j}c_{\vec{\sigma}}c^*_{\vec{1}-\vec{\sigma}}.
\end{equation}
Interchanging the order of summation and applying the identity Eq. \eqref{identity} again yields
\begin{equation}
\sum_{\{\vec{a}:\sum_{j=1}^N a_j \, \mathrm{even}\}} \sum_{\vec{\sigma}\in \{0,1\}^N} e^{i\pi \sum_{j=1}^N \sigma_j a_j}c_{\vec{\sigma}}c^*_{\vec{1}-\vec{\sigma}} = \sum_{\vec{\sigma}\in \{0,1\}^N} c_{\vec{\sigma}}c^*_{1-\vec{\sigma}} 2^{N-1}(\delta_{\vec{\sigma}\vec{0}} + \delta_{\vec{\sigma}\vec{1}}) = 2^{N-1} (c_{\vec{0}}c_{\vec{1}}^* + c_{\vec{1}}c_{\vec{0}}^*).
\end{equation}
and we thus arrive at the compact expression
\begin{equation}
\label{eq:appwinprob}
p_{\mathrm{qu}}(|\psi\rangle) = \frac{1}{2}\left(1+ c_{\vec{0}}c_{\vec{1}}^* + c_{\vec{1}}c_{\vec{0}}^*\right).
\end{equation}
It remains to establish the connection with the even and odd parity GHZ states $|\mathrm{GHZ}^{\pm}\rangle = \frac{1}{\sqrt{2}}(|\vec{0}\rangle \pm |\vec{1}\rangle)$. To this end, note that
\begin{equation}
|\langle \psi | \mathrm{GHZ}^{\pm} \rangle|^2 = \frac{1}{2}|c_{\vec{0}} \pm c_{\vec{1}}|^2,
\end{equation}
which implies that
\begin{equation}
|\langle \psi | \mathrm{GHZ}^+ \rangle|^2-|\langle \psi | \mathrm{GHZ}^- \rangle|^2 = \frac{1}{2}(|c_{\vec{0}}+c_{\vec{1}}|^2-|c_{\vec{0}}-c_{\vec{1}}|^2) = c_{\vec{0}}c_{\vec{1}}^*+c_{\vec{1}}c_{\vec{0}}^*.
\end{equation}
We deduce finally that Eq. \eqref{eq:appwinprob} can be written as
\begin{equation}
\label{eq:appthm}
p_{\mathrm{qu}}(|\psi\rangle) = \frac{1}{2}\left(1+ |\langle \psi | \mathrm{GHZ}^+\rangle |^2 -|\langle \psi | \mathrm{GHZ}^-\rangle |^2\right),
\end{equation}
which was to be shown.
\section{Boyer's modulo $M$ generalization of the parity game}
\label{app:Boyer}

In this Appendix, we extend various ideas from the main text to Boyer's ``modulo $M$'' generalization of the parity game\cite{boyer2004extended}. This is an $N$-player nonlocal game which can be won with certainty using the qudit GHZ state
\begin{equation}
|\mathrm{GHZ}_M\rangle = \frac{1}{\sqrt{M}}(|0\rangle^{\otimes N} + |1\rangle^{\otimes N} + \ldots + |M-1\rangle^{\otimes N}).
\end{equation}

The rules of Boyer's modulo $M$ game are modelled on the Brassard-Broadbent-Tapp rules for the parity game\cite{brassard2005recasting}. The game is determined by two positive integers, the \emph{divisor} $D$ and the \emph{modulus} $M$. There are $N$ players and player $j$ receives a number $a_j \in \{0,1,\ldots,D-1\}$. The promise is that $D$ divides $\sum_{j=1}^N a_j$ and to win the game, player $j$ must return $b_j \in \{0,1,\ldots,M-1\}$ such that
\begin{equation}
\sum_{j=1}^N b_j \equiv \frac{\sum_{j=1}^N a_j}{D} \,\mathrm{mod}\,{M}.
\end{equation}

For all $D$ and $M$, the modulo $M$ game admits a perfect quantum strategy $\mathcal{S}_{\mathrm{Boyer}} = (|\mathrm{GHZ}_M\rangle,\mathcal{P}_{\mathrm{Boyer}})$. In what follows, it will be useful to define single qudit gates
\begin{equation}
\hat{C}|k \rangle = \omega_{M}^{k} |k\rangle, \quad \hat{S}|k\rangle = |k+1\rangle,\quad \hat{W}|k\rangle = \frac{1}{\sqrt{M}} \left(|0\rangle + \omega_M^k |1\rangle +\ldots \omega_M^{k(M-1)} |M-1\rangle\right)
\end{equation}
where $\omega_M = e^{i 2 \pi/M}$. These correspond to a ``clock'' operator, a ``shift'' operator, and a change of basis from clock to shift eigenstates respectively. 

The Boyer protocol $\mathcal{P}_{\mathrm{Boyer}}$, which always wins the modulo $M$ game if the players share the state $|\mathrm{GHZ}_M\rangle$ before playing, consists of the following steps:
\begin{enumerate}
    \item Each player acts with $\hat{C}^{-a_j/D}$ on their qudit.
    \item Each player acts with $\hat{W}$ on their qudit.
    \item Each player measures their qudit in the clock basis and returns the qudit value $b_j$.
\end{enumerate}
Let us briefly summarize why this works. After the first step, the shared GHZ state is given by
\begin{equation}
|\psi'\rangle = \frac{1}{\sqrt{M}} \left(|0 \rangle^{\otimes N} + \omega_M^{-\sum_{j=1}^N a_j/D}|1\rangle^{\otimes N} + \ldots + \omega_M^{-(M-1)\sum_{j=1}^N a_j/D}|1\rangle^{\otimes N} \right)
\end{equation}
After the second step, this becomes
\begin{align}
\nonumber |\psi'' \rangle &= \sum_{k=0}^{M-1} \frac{1}{\sqrt{M}} \omega_M^{-k \sum_{j=1}^N a_j/D} \bigotimes_{j=1}^N \left(\frac{1}{\sqrt{M}}  \sum_{y=0}^{M-1} \omega_M^{k y} |y\rangle\right) \\
\nonumber &= \frac{1}{M^{(N-1)/2}}  \sum_{\vec{y} \in \{0,1,\ldots,M-1\}^N} \left(\frac{1}{M} \sum_{k=0}^{M-1} \omega_M^{k \sum_{j=1}^N (y_j-a_j/D)}\right) |y_1y_2\ldots y_N\rangle \\
&=  \frac{1}{M^{(N-1)/2}} \sum_{\substack{\vec{y} \in \{0,1,\ldots,M-1\}^N\\ \sum_{j=1}^N y_j \equiv \sum_{j=1}^N a_j/D \,\mathrm{mod}\,{M}}} |y_1 y_2 \ldots y_M\rangle.
\end{align}
From here it is clear that a measurement in the clock basis wins the game. 

In relating the parity game to condensed matter systems in the main text, we used the fact that the qubit GHZ state can be interpreted as the ground state of a quantum Ising model. The qudit GHZ state can similarly be interpreted as a ground state of a quantum clock model. A generic quantum clock model has the form
\begin{equation}
\label{eq:clock}
\hat{H} = -J\sum_{j=1}^N \frac{1}{2}(\hat{C}_j^\dagger \hat{C}_{j+1} +\hat{C}_{j+1}^\dagger \hat{C}_{j}) - \Gamma \sum_{j=1}^N \frac{1}{2}(\hat{S}_j + \hat{S}_j^\dagger) - h \sum_{j=1}^N \frac{1}{2}(\hat{C}_j + \hat{C}_j^\dagger)
\end{equation}
where the physical meaning of the various coefficients is much the same as for the corresponding coefficients in the quantum Ising model Eq. \eqref{eq:LTFIM} and we again set $J, \Gamma >0$ and assume periodic boundary conditions. In the limit $h=0, \, \Gamma \to 0^+$, degenerate perturbation theory implies that the ground state of this model is given by $|\mathrm{GHZ}_M\rangle$. A generalization of the problem we considered in the main text to qudits is to consider the probability of success for the quantum strategy $\mathcal{S} = (|\psi_0\rangle,\mathcal{P}_{\mathrm{Boyer}})$ for the modulo $M$ game with divisor $M$, where $|\psi_0\rangle$ denotes the ground state of the clock model Eq. \eqref{eq:clock}. By our results for the quantum Ising model, we expect that
a $\mathbb{Z}_M$-symmetry-breaking field $h \neq 0$ will eliminate the quantum advantage of $\mathcal{S}$, while if $h=0$, quantum advantage of $\mathcal{S}$ will persist over some non-zero range of $\Gamma>0$ as $N\to \infty$.

We now report an analogue of our Theorem \ref{thm:thm} for Boyer's modulo $M$ game. Specifically, we have the following result:
\begin{theorem}
\label{thm:thm2}
The quantum strategy $\mathcal{S} = (|\psi\rangle,\mathcal{P}_{\mathrm{Boyer}})$ wins the modulo $M$ game with probability
\begin{equation}
\label{eq:quditfidelity}
p_{\mathrm{qu}}(|\psi\rangle) = \frac{1}{M} \left(1 + \sum_{l=1}^{M-1} \sum_{\vec{y} \in \{0,1,\ldots,M-1-l\}^N} |\langle \psi | \varphi^+_{\vec{y},\vec{y}+\vec{l}}\rangle|^2- |\langle \psi | \varphi^-_{\vec{y},\vec{y}+\vec{l}}\rangle|^2\right),
\end{equation}
where we defined the family of $N$-qudit cat states
\begin{equation}
\label{eq:zmcat}
|\varphi^{\pm}_{\vec{y},\vec{z}}\rangle = \frac{1}{\sqrt{2}}(|y_1 y_2\ldots y_N\rangle \pm |z_1 z_2\ldots z_N\rangle).
\end{equation}
\end{theorem}
This result is more complicated than the corresponding expression Eq. \eqref{eq:appthm} for the parity game; this more complicated structure reflects the multiple distinct ways of realizing GHZ-state-like multipartite entanglement within a system of many qudits.

We next prove that the qudit GHZ state $|\mathrm{GHZ}_M\rangle$ is the unique pure state $|\psi\rangle$ such that the quantum strategy $\mathcal{S} = (|\psi\rangle, \mathcal{P}_{\mathrm{Boyer}})$ wins the modulo $M$ game with certainty. To achieve this, it is helpful to write the (normalized) state $|\psi\rangle$ in question explicitly as
\begin{equation}
|\psi\rangle = \sum_{\vec{y} \in \{0,1,\ldots,M-1\}^N} c_{\vec{y}} |y_1 y_2\ldots y_N\rangle
\end{equation}
in the qudit computational basis. In components, the fidelity formula Eq. \eqref{eq:zmcat} reads
\begin{equation}
p_{\mathrm{qu}}(|\psi\rangle) = \frac{1}{M}\left(1 + \sum_{l=1}^{M-1} \sum_{\vec{y} \in \{0,1,\ldots,M-1-l\}^N} c_{\vec{y}}c_{\vec{y}+\vec{l}}^* + c_{\vec{y}+\vec{l}}c_{\vec{y}}^*\right).
\end{equation}
To proceed further, we must rewrite this as a sum over mutually orthogonal subspaces of the $N$-qudit Hilbert space. The subspaces in question can be labelled by multi-indices $\vec{y}$ with $\min_j y_j = 0$, and are each spanned by sets of computational basis vectors related to one another by actions of the global shift operator $\hat{S} = \prod_{i=1}^L \hat{S}_i$ and its inverse. Reordering the summation to reflect this structure, we obtain
\begin{align}
\nonumber p_{\mathrm{qu}}(|\psi\rangle) &= \frac{1}{M}\left(1 + \sum_{\substack{\vec{y} \in \{0,1,\ldots,M-2\}^N \\ \min_j y_j = 0}} \sum_{l=1}^{M-1-\max_j y_j} c_{\vec{y}}c_{\vec{y}+\vec{l}}^* + c_{\vec{y}+\vec{l}}c_{\vec{y}}^*\right) \\
&= \frac{1}{M}\left(1 + \sum_{\substack{\vec{y} \in \{0,1,\ldots,M-2\}^N \\ \min_j y_j = 0}} \left|\sum_{l=0}^{M-1-\max_j y_j} c_{\vec{y}+\vec{l}}\right|^2 - \sum_{l=0}^{M-1-\max_j y_j} |c_{\vec{y}+\vec{l}}|^2\right).
\end{align}
By normalization $\sum_{\vec{y}\in \{0,1,\ldots,M-1\}^N} |c_{\vec{y}}|^2 = 1$, this can be written as
\begin{equation}
p_{\mathrm{qu}}(|\psi\rangle) = \frac{1}{M}\left(\sum_{\substack{\vec{y} \in \{0,1,\ldots,M-1\}^N \\ \min_j y_j = 0, \,\max_j y_j = M-1}}| c_{\vec{y}}|^2 + \sum_{\substack{\vec{y} \in \{0,1,\ldots,M-1\}^N \\ \min_j y_j = 0,\, \max_j y_j < M-1}} \left|\sum_{l=0}^{M-1-\max_j y_j} c_{\vec{y}+\vec{l}}\right|^2\right),
\end{equation}
which is the desired sum over mutually orthogonal subspaces and defines a block-diagonal Hermitian form $\hat{A}$ via the relation $p_{\mathrm{qu}}(|\psi \rangle) = \langle \psi | \hat{A}| \psi \rangle$. Maximizing this Hermitian form subject to the normalization constraint $\langle \psi | \psi \rangle=1$, it follows that $p_{\mathrm{qu}}({|\psi\rangle})$ attains its maximum $\lambda^*$ when $|\psi\rangle = |v^*\rangle$ is an eigenvector of $\hat{A}$ with the largest eigenvalue $\lambda^*$ among all such eigenvectors. In particular, the quantum strategy $\mathcal{S} = (|\psi\rangle, \mathcal{P}_{\mathrm{Boyer}})$ is a perfect quantum strategy for the modulo $M$ game if and only if $|\psi\rangle$ is $\hat{A}$-invariant, with $\hat{A} |\psi\rangle = |\psi\rangle$.

It remains to obtain the spectrum of $\hat{A}$. First we label each block of $\hat{A}$ by a multi-index $\vec{y}\in\mathbb{Z}_M^N$ with $\min_j y_j = 0$, and note that each block in this subspace has matrix elements
\begin{equation}
B_{ll'} = A_{\vec{y}+\vec{l},\vec{y}+\vec{l'}} = \frac{1}{M}, \quad l = 0, 1, \ldots M-1-\max_j y_j.
\end{equation}
By the Cauchy-Schwartz inequality, any unit norm vector in this subspace satisfies
\begin{equation}
\langle v |\hat{B} | v\rangle = \sum_{l,l'=0}^{M-1-\max_j y_j} v_l^*B_{ll'}v_{l'} =\frac{1}{M}\left|\sum_{l=0}^{M-1-\max_j y_j} v_{l}\right|^2 \leq \frac{M-\max_{j} y_j}{M},
\end{equation}
which is saturated if and only if $v_l = e^{i\varphi}/\sqrt{M-\max_{j} y_j}$, where $\varphi$ is an arbitrary global phase. Thus the largest eigenvalue in this block is $\lambda^*_{\vec{y}} = (M-\max_j y_j)/M$, and its eigenspace is spanned by $v_l = 1/\sqrt{M-\max_{j} y_j}$.

It follows that the largest eigenvalue of $\hat{A}$ is equal to $\lambda^*_{\vec{0}} = 1$ and resides in the unique block indexed by the multi-index $\vec{y} = (0,0,\ldots,0)$ and consisting of the states $\{|0\rangle^{\otimes N}, \, |1 \rangle^{\otimes N}, \, \ldots |M-1\rangle^{\otimes N}$\}. Furthermore, the eigenspace corresponding to this eigenvalue is spanned by
\begin{equation}
c_{\vec{y}} = \begin{cases} 0 & \min_{j} y_j \neq \max_j y_j \\ \frac{1}{\sqrt{M}} & \min_{j} y_j = \max_j y_j\end{cases},
\end{equation}
which precisely the qudit GHZ state $|\mathrm{GHZ}_M\rangle$. We deduce that $p_{\mathrm{qu}}(|\psi\rangle)$ attains its maximum value of 1 if and only if $|\psi\rangle$ is equal to the state $|\mathrm{GHZ}_M\rangle$, up to a global phase. 

An interesting corollary of our result is that applying Boyer's protocol to an $M'$ qudit GHZ state embedded suitably in the $M$ qudit Hilbert space, where $2 \leq M' <M$ is smaller than $M$, still confers some quantum advantage over random guessing for the modulo $M$ game, with
\begin{equation}
p_{\mathrm{qu}}(|\mathrm{GHZ}_{M'}\rangle) = \frac{M'}{M} > \frac{1}{M}.
\end{equation}

Finally, we note that the optimal classical strategy for this game is not known in general. The initial work\cite{boyer2004extended} proved that provided $M$ and $D$ are not coprime, the game cannot always be won classically, but did not estimate the optimal classical probability of winning for general values of $M$. By adapting the arguments of Brassard-Broadbent-Tapp\cite{brassard2005recasting} and Mermin\cite{mermin1990extreme}, we have found that the optimal classical probability of winning is bounded above by $1/M$ plus a correction that is exponential in $N$. To prove this, let $b_j(a_j)$ be a classical deterministic strategy, whereby player $j$ always returns the output $b_j(a_j) \in \{0,1,\ldots,M-1\}$ given the input $a_j \in \{0,1,\ldots,D-1\}$. Let us write
\begin{equation}
S_{j}(a_j) = \omega_M^{b_j(a_j)}
\end{equation}
and also introduce the $D$th root of unity $\omega_D = e^{i2\pi/D}$. We then consider the quantities
\begin{align}
\nonumber \lambda_n &= \frac{1}{D} \sum_{k=0}^{D-1} \prod_{j=1}^N \left[S_{j}(0)^n + \omega_D^{k-n/M}S_{j}(1)^n + \ldots + \omega_D^{(k-n/M)(D-1)}S_{j}(D-1)^n \right] \\
\nonumber &= \frac{1}{D} \sum_{k=0}^{D-1} \sum_{\vec{a} \in \{0,1,\ldots,D-1\}^N} \omega_D^{\sum_{j=1}^N (k-n/M)a_j} \prod_{j=1}^N S_{j}(a_j)^n \\
\nonumber &=\sum_{\vec{a} \in \{0,1,\ldots,D-1\}^N} \omega_D^{-n/M\sum_{j=1}^N a_j} \left(\frac{1}{D} \sum_{k=0}^{D-1}  \omega_D^{k \sum_{j=1}^N a_j}\right) \prod_{j=1}^N S_{j}(a_j)^n \\
\nonumber &=\sum_{\substack{\vec{a} \in \{0,1,\ldots,D-1\}^N \\ D | \sum_{j=1}^N a_j }} \omega_M^{-n/D\sum_{j=1}^N a_j } \prod_{j=1}^N S_{j}(a_j)^n \\
\nonumber &= \sum_{\substack{\vec{a} \in \{0,1,\ldots,D-1\}^N \\ D | \sum_{j=1}^N a_j }} \omega_M^{n \left(\sum_{j=1}^N b_j -\sum_{j=1}^N a_j/D\right)}\\
&= N_{\mathrm{wins}} + N_{1} \omega^n_M + \ldots + N_{M-1} \omega_M^{n(M-1)}
\end{align}
where $n=0,1,\ldots,M-1$, $N_{\mathrm{wins}}$ denotes the number of winning inputs to this strategy and $N_{r}$ the number of inputs that this strategy loses by a margin $r \equiv \sum_{j=1}^N b_j -\sum_{j=1}^N a_j/D \,\mathrm{mod}\, M$. We also let
\begin{equation}
N_{\mathrm{losses}} = \sum_{r=1}^{M-1} N_r
\end{equation}
denote the total number of losing inputs to this strategy. It is clear that
\begin{equation}
\lambda_0 = N_{\mathrm{wins}} + N_{\mathrm{losses}} = D^{N-1}
\end{equation}
while for $1 \leq n \leq M-1$,
\begin{equation}
\lambda_n \leq |\lambda_n| \leq \left(\max_{k,b_j} \left|\omega_M^{nb_j(0)} + \omega_D^{k-n/M}\omega_M^{nb_j(1)} + \ldots + \omega_D^{(k-n/M)(D-1)}\omega_M^{nb_j(D-1)}\right|\right)^N.
\end{equation}
In general these inequalities are not saturated, so the bound is not tight. It is convenient to write
\begin{equation}
\label{eq:ub}
s_{D,M} = \max_{n,k,b} \left|\omega_M^{nb(0)} + \omega_D^{k-n/M}\omega_M^{nb(1)} + \ldots + \omega_D^{(k-n/M)(D-1)}\omega_M^{nb(D-1)}\right|,
\end{equation}
where the maximization is over all $n=1,2,\ldots,M-1$, $k=0,1,\ldots,D-1$ and functions $b : \mathbb{Z}_D \to \mathbb{Z}_M$. Then, since
\begin{equation}
\frac{1}{M}\sum_{n=0}^{M-1} \lambda_n = \frac{1}{M}\sum_{n=0}^{M-1} \left(N_{\mathrm{wins}} + N_{1} \omega^n_M + \ldots + N_{M-1} \omega_M^{n(M-1)}\right) = N_{\mathrm{wins}},
\end{equation}
it follows that
\begin{equation}
N_{\mathrm{wins}} \leq \frac{1}{M}\left(D^{N-1} + (M-1)\max_{1\leq n \leq M-1} |\lambda_n|\right) \leq \frac{1}{M}\left(D^{N-1} + (M-1)s_{D,M}^N\right)
\end{equation}
and in particular that the classical winning probability
\begin{equation}
p_{\mathrm{cl}}(D,M) = \frac{N_{\mathrm{wins}}}{D^{N-1}} \leq \frac{1}{M} + \frac{M-1}{M} s_{D,M} \left(\frac{s_{D,M}}{D}\right)^{N-1}
\end{equation}
for any given strategy. Thus the optimal classical winning probabilty
\begin{equation}
p^*_{\mathrm{cl}}(D,M) \leq \frac{1}{M} + \frac{M-1}{M} s_{D,M} \left(\frac{s_{D,M}}{D}\right)^{N-1}.
\end{equation}

By enumerating all $(M-1)DM^D$ cases in Eq. \eqref{eq:ub} numerically, we find that
\begin{equation}
s_{3,3} \approx 2.53 < 3, \quad s_{4,4} \approx 3.62 < 4, \quad s_{5,5} \approx 4.69 < 5, \ldots
\end{equation}
In general, we expect that $s_{M,M}<M$, since Eq. \eqref{eq:ub} involves the root of unity $\omega_{M^2}$ and thus cannot attain its maximum value for the allowed values of $n$. This implies that when $D=M$, the probability of winning for the optimal classical strategy is constrained to equal $1/M$ as $N \to \infty$, i.e. performs no better than random guessing in the large-system limit. To our knowledge, this upper bound is a substantial improvement on existing results\cite{boyer2004extended} for $p^*_{\mathrm{cl}}(D,M)$ for generic values of $M$. However, this type of reasoning only seems to yield a tight upper bound for the original parity game with $D=M=2$; finding a tight bound (and thereby determining the optimal classical strategy) for general values of $D$ and $M$ remains an open question.

\end{document}